\documentclass[onecolumn,useAMS,usenatbib]{mn2e}
\usepackage{graphicx}
\usepackage{natbib}

\usepackage{bm}
\usepackage{amsfonts}
\usepackage{latexsym}
\usepackage[latin1]{inputenc}
\usepackage{amsmath}
\usepackage{epsfig}
\usepackage{amsbsy}
\usepackage{mnfixes}

\newcommand{\mtb}[1]{\mathbf{#1}}

\newcommand{\n}{\mathrm n}
\newcommand{\p}{\mathrm p}

\newcommand{\x}{\mathrm x}
\newcommand{\y}{\mathrm y}
\newcommand{\s}{\mathrm s}
\newcommand{\ord}{\mathrm o}

\newcommand{\veps}{\varepsilon}

\def \veps{\varepsilon}
\def\jnl@style{\it}
\def\aaref@jnl#1{{\jnl@style#1}}
\def\aaref@jnl#1{{\jnl@style#1}}

\def\aj{\aaref@jnl{AJ}}                   % Astronomical Journal
\def\apj{\aaref@jnl{ApJ}}                 % Astrophysical Journal
\def\apjl{\aaref@jnl{ApJ}}                % Astrophysical Journal, Letters
\def\apjs{\aaref@jnl{ApJS}}               % Astrophysical Journal, Supplement
\def\apss{\aaref@jnl{Ap\&SS}}             % Astrophysics and Space Science
\def\aap{\aaref@jnl{A\&A}}                % Astronomy and Astrophysics
\def\aapr{\aaref@jnl{A\&A~Rev.}}          % Astronomy and Astrophysics Reviews
\def\aaps{\aaref@jnl{A\&AS}}              % Astronomy and Astrophysics, Supplement
\def\mnras{\aaref@jnl{MNRAS}}             % Monthly Notices of the RAS
\def\prc{\aaref@jnl{Phys.~Rev.~C}}        % Physical Review C
\def\prd{\aaref@jnl{Phys.~Rev.~D}}        % Physical Review D
\def\prl{\aaref@jnl{Phys.~Rev.~Lett.}}    % Physical Review Letters
\def\qjras{\aaref@jnl{QJRAS}}             % Quarterly Journal of the RAS
\def\skytel{\aaref@jnl{S\&T}}             % Sky and Telescope
\def\ssr{\aaref@jnl{Space~Sci.~Rev.}}     % Space Science Reviews
\def\zap{\aaref@jnl{ZAp}}                 % Zeitschrift fuer Astrophysik
\def\nat{\aaref@jnl{Nature}}              % Nature
\def\aplett{\aaref@jnl{Astrophys.~Lett.}} % Astrophysics Letters
\def\apspr{\aaref@jnl{Astrophys.~Space~Phys.~Res.}} % Astrophysics Space Physics Research
\def\physrep{\aaref@jnl{Phys.~Rep.}}      % Physics Reports
\def\physscr{\aaref@jnl{Phys.~Scr}}       % Physica Scripta

\title{Oscillations of Rapidly Rotating Superfluid Stars}

\author[A. Passamonti et al]
{A. Passamonti\thanks{E-mail:a.passamonti@soton.ac.uk} , B. Haskell,
N. Andersson \\ \\
\\ \\
School of Mathematics, University of Southampton, Southampton SO17 1BJ, UK}

\begin{document}

%%%%%%%%%%%%%%%%%%%%%%%%%%%%%%%%%%%%  DATE  %%%%%%%%%%%%%%%%%%%%%%%%%%%%%%%%%%%%
\date{\today}

%%%%%%%%%%%%%%%%%%%%%%%%%%%%%%% PAGE RANGE  %%%%%%%%%%%%%%%%%%%%%%%%%%%%%%%%%%%%
\pagerange{\pageref{firstpage}--\pageref{lastpage}} \pubyear{}

%%%%%%%%%%%%%%%%%%%%%%%%%%%%%  MAKETITLE  %%%%%%%%%%%%%%%%%%%%%%%%%%%%%%%%%%%%%%
\maketitle

%%%%%%%%%%%%%%%%%%%%%%%%%%%%%%% FIRST PAGE  %%%%%%%%%%%%%%%%%%%%%%%%%%%%%%%%%%%%
\label{firstpage}

%%%%%%%%%%%%%%%%%%%%%%%%%%%%%  ABSTRACT  %%%%%%%%%%%%%%%%%%%%%%%%%%%%%%%%%%%%%%%

\begin{abstract}

Using time evolutions of the relevant linearised equations we study
non-axisymmetric oscillations of rapidly rotating and superfluid
neutron stars. We consider perturbations of Newtonian axisymmetric
background configurations and account for the presence of superfluid
components via the standard two-fluid model.  Within the Cowling
approximation, we are able to carry out evolutions for uniformly
rotating stars up to the mass-shedding limit. This leads to the first
detailed analysis of superfluid neutron star oscillations in the fast
rotation regime, where the star is significantly deformed by the
centrifugal force. For simplicity, we focus on background models where
the two fluids (superfluid neutrons and protons) co-rotate, are in
$\beta$-equilibrium and coexist throughout the volume of the star. We
construct sequences of rotating stars for two analytical model
equations of state. These models represent relatively simple
generalisations of single fluid, polytropic stars. We study the
effects of entrainment, rotation and symmetry energy on non-radial
oscillations of these models. Our results show that entrainment and
symmetry energy can have a significant effect on the rotational
splitting of non-axisymmetric modes. In particular, the symmetry
energy modifies the inertial mode frequencies considerably in the
regime of fast rotation.

\end{abstract}

%%%%%%%%%%%%%%%%%%%%%%%%%%%%%  PACS  %%%%%%%%%%%%%%%%%%%%%%%%%%%%%%%%%%%%%%%%%%%
%\pacs{04.30.Db, 04.40.Dg, 95.30.Sf, 97.10.Sj}
%%%%%%%%%%%%%%%%%%%%%%%%%%%%%  Keywords  %%%%%%%%%%%%%%%%%%%%%%%%%%%%%%%%%%%%%%%%%%%

%%%%%%%%%%%%%%%%%%%%%%%%%%%%%%% SEC. %%%%%%%%%%%%%%%%%%%%%%%%%%%%%%%%%%%%%%%%%%%
\section{Introduction} \label{sec:Intro}
%%%%%%%%%%%%%%%%%%%%%%%%%%%%%%%%%%%%%%%%%%%%%%%%%%%%%%%%%%%%%%%%%%%%%%%%%%%%%%%%

According to the standard paradigm, millisecond pulsars are
accelerated to their fast rotation rates by accreting matter from a
close companion. This means that they tend to be relatively
old. Moreover, the fastest spinning pulsars should have weak
(exterior) magnetic fields.  In the standard accretion model, neutron
stars with canonical $10^{12}$~G dipole fields will reach equilibrium
already at a modest spin. A weak surface field is also expected since
accretion leads to magnetic field burial. This picture agrees well
with observational data.  Rapidly rotating neutron stars are most
commonly found in binary systems. It is well established that
accreting neutron stars in low-mass X-ray binaries (where the angular
momentum transfer is more efficient due to the long evolution time of
the low mass partner) can reach a millisecond rotation
period. Furthermore, the fastest known millisecond pulsar
J1748-2446ad, with a period of 1.39~ms~\citep{2006Sci...311.1901H}, is
in a binary system. In fact, its companion, with mass $M \geq 0.14
M_{\odot}$, could still fill the Roche lobe powering the spin-up phase
further. The mass and radius of J1748-2446ad are unknown, but
combining reasonable ranges for these parameters, $M = 1.4-2
M_{\odot}$ and $R=10-14~\rm{km}$, with an empirical formula for the
maximum rotation of the star~\citep{2004Sci...304..536L}
\begin{equation}
\Omega_K \approx 6566 \left( \frac{M}{M_{\odot}} \right) ^{1/2}
\left( \frac{ 10 \rm{km} }{ R } \right) ^{3/2}  \rm{Hz} \, ,
\end{equation}
one finds that the spin of this object lies in the range $0.48
\lesssim \Omega / \Omega_K \lesssim 0.96 $.  In other words, it could
be close to the mass-shedding limit.

The temperature of a mature neutron star is likely below the critical
temperature, $T_c \simeq 5\times 10 ^9$~K, where neutrons and protons
are superfluid and superconducting, respectively. Depending on the
cooling mechanism, neutrino emission can cool a hot proto-neutron star
below this temperature shortly after its formation in a core collapse
supernova~\citep*[see e.g,][]{2006NuPhA.777..497P}.  In an accreting
system, the neutron star core temperature is not expected to increase
beyond $T_c$ (nuclear burning in the accreted surface layers is
thought to the heat the core to $\sim 10^8$~K).  Hence, all mature
neutron stars should contain degenerate superfluid neutrons in the
outer core and the inner crust and degenerate superconducting protons
in the outer core. The deep core may
contain more exotic phases of matter, like superfluid hyperons and/or
colour superconducting quarks.  Superfluidity influences the thermal
evolution and the dynamical properties of a neutron star. In
particular, the dynamics is strongly affected by entrainment, the
formation of quantized neutron vortices, and the presence of new
dissipative mechanisms like mutual friction.  An understanding of
superfluid dynamics is crucial for modelling many aspects of the
neutron star physics, ranging from pulsar glitches and free precession
to the mutual friction damping of stellar oscillations and associated
instabilities.

Tidal forces, accretion and glitches may trigger oscillations and/or
instabilities in rapidly rotating neutron stars. Observations of such
oscillations, either via electromagnetic or gravitational radiation,
would help us explore the exotic physics of these compact objects
\citep{1998MNRAS.299.1059A,benhar-2004-70,2007MNRAS.374..256S}.  In
this context it is interesting to note the differences in dynamics
between neutron stars above the superfluid transition temperature and
colder systems.  The core of a hot young neutron star is relatively
well described by the Navier-Stokes equations.  In contrast, a star
with a superfluid core requires a multi-fluid description. The
standard model for such systems is inspired by the two-fluid model for
superfluid Helium~\citep{Landau:1959, Khalatnikov:1965,
Tilley:1990}. The oscillation spectrum of superfluid neutron stars
reflects the presence of the additional degree of
freedom~\citep{1988ApJ...333..880E}. Basically, the perturbed fluid
elements in a two-fluid system can oscillate either in phase or in a
counter-phase motion. Previous studies, see for example, \citet{1995A&A...303..515L,
2001MNRAS.328.1129A, 2002A&A...393..949P, 2003MNRAS.344..207Y}, have
established that co-moving pulsations have spectral properties similar
to single fluid stars. Hence, it is natural to refer to such modes as
``ordinary'' modes. The counter-moving degree of freedom leads to new
oscillation modes that are specific to the two-fluid systems. These
are often referred to as ``superfluid'' modes. Later, when we write
down the perturbation equations for a superfluid neutron star core, we
choose to work with variables that are directly associated with the
two types of motion. This is natural if we want to distinguish
spectral properties associated with the ``superfluid" degree of
freedom. In addition, we use the standard classification of neutron
star oscillation modes, based on the main restoring force that acts on
a perturbed fluid element. A rotating single fluid star can sustain
acoustic and inertial modes restored by pressure and the Coriolis
force, respectively.  When thermal or composition gradients are
present in the star, buoyancy acts as restoring force for the
so-called gravity modes~\citep{1989nos..book.....U,
1992ApJ...395..240R}. Previous work has shown that superfluid neutron
stars have two families of acoustic and inertial modes, more or less
clearly (depending on the stellar model) associated with the co- and
counter-moving fluid motion. It is, however, not the case that all
single fluid modes have a "double" in the superfluid problem. The
gravity modes are not present at all in a superfluid
core~\citep{1995A&A...303..515L, 2001MNRAS.328.1129A,
2002A&A...393..949P}.  Their absence provides a potentially important
signature for neutron star seismology.

Rapidly rotating neutron stars have been studied in detail with a
variety of methods~\citep[see e.g.,][]{lrr-2003-3}. Yet, there have
not been any previous studies of multi-fluid dynamics in the rapid
rotation regime near the break-up limit.  The oscillation modes of
superfluid neutron stars have only been calculated in the frequency
domain using the slow rotation
approximation~\citep{2000PhRvD..61j4003L, 2003MNRAS.344..207Y,
2003PhRvD..67l4019Y}.  In that framework, the effects of the stellar
rotation are determined perturbatively as corrections to the
non-rotating results.  The only previous attempt (that we are aware
of) to study superfluid oscillations for truly fast spinning stars is
the work by \citet{2000PhRvD..61j4003L}.  They extended the
two-potential formalism of \citet{1990ApJ...355..226I} to the
superfluid case.  However, due to numerical difficulties they could
not study rotating models near the break-up limit.  Neither did they
manage to determine the superfluid modes.

In this work, we study the time evolution of perturbed fast rotating,
Newtonian superfluid neutron stars within the Cowling approximation.
As far as we are aware, this is the first study that evolves in time
the oscillations of superfluid neutron stars. Moreover, it is the
first detailed analysis of the rapid rotation regime.  Within the
framework of the two-fluid formalism, we carry out a linear
perturbation analysis for stationary and axisymmetric equilibrium
configurations. As preparation for more detailed studies, we consider
relatively simple models where the two fluids coexist throughout the
star, and where the unperturbed configuration is in
$\beta$-equilibrium.  These assumptions imply that the two fluids are
co-rotating and share the same external stellar surface in our
background configurations. We use these models to investigate the
effects of entrainment, symmetry energy and rotation on the superfluid
oscillation spectrum. In order to establish the reliability of our
numerical evolution code, we compare our results to previous work for
non-rotating and slowly rotating models. We then consider, for the
first time, the dynamics of superfluid models rotating up to the mass
shedding limit.

%%%%%%%%%%%%%%%%%%%%%%%%%%%%%%% SEC. %%%%%%%%%%%%%%%%%%%%%%%%%%%%%%%%%%%%%%%%%%%
\section{Equations of Motion} \label{sec:pert-eqs}
%%%%%%%%%%%%%%%%%%%%%%%%%%%%%%%%%%%%%%%%%%%%%%%%%%%%%%%%%%%%%%%%%%%%%%%%%%%%%%%%

We use the two-fluid framework for superfluid neutron
stars~\citep{1991ApJ...380..515M, 1991ApJ...380..530M,
2004PhRvD..69d3001P, 2006CQGra..23.5505A}. This model distinguishes
between a superfluid neutron component and a neutral mixture of
protons and electrons.  The charged particles are assumed to be locked
together by the electromagnetic interaction on a timescale much shorter
than the dynamics we consider. For simplicity, we refer to the charged
particle conglomerate as the ``protons'' from now on.
When the mass of each component is conserved the dynamics is described
by two mass conservation laws, two Euler-type equations and the
Poisson equation for the gravitational
potential~\citep{2004PhRvD..69d3001P}:
\begin{equation}
\partial_t \rho_{\x} + \nabla_{i} \left( \rho_{\x} v_{\x}^{i} \right) = 0  \, ,  \label{eq:Mcon}\\
\end{equation}
\begin{equation}
\left( \partial_t + v_{\x}^{k} \nabla_{k} \right)
 \left( v_{i}^{\x} + \varepsilon_{\x} w_{i}^{\y\x} \right) + \nabla_{i} \left( \Phi + \tilde{\mu}_{\x} \right)
+ \varepsilon_{\x} w_{k}^{\y\x} \nabla_{i} v_{\x}^{k} =  \frac{f^{\x}_{i}}{\rho_{\x}} \, ,  \label{eq:Euler} \\
\end{equation}
\begin{equation}
 \nabla^2 \Phi = 4 \pi G \rho \, . \label{eq:Poisson}
\end{equation}
Here the indices $i$ and $k$ label the spatial components of the
vectors while $\x$ and $\y$ denote the two fluid components. The
constituent indices take the values $\n$ for the neutrons and $\p$ for
the protons. Note that the summation rule for repeated indices applies
only for spatial and not for constituent indices. In
equations~(\ref{eq:Mcon})--(\ref{eq:Poisson}), $\rho = \rho_\n +
\rho_\p$ and $\Phi$ represent the total mass density and the
gravitational potential, respectively. Meanwhile, $\mtb{f}^{\x}$ is
the force density acting on the $\x$ fluid component. In this work, we
consider an isolated system where dissipation processes, like mutual
friction, are absent. We then have $\mtb{f}^{\x}=\mtb{0}$.
Furthermore, we have assumed that the particle masses are equal, $m=m_{\n}=m_{\p}$, and defined the
chemical potential and the relative velocity between the two
fluids as follows:
\begin{eqnarray}
\tilde{\mu}_{\x} & \equiv & \left. \frac{\partial \mathcal{E}}{\partial \rho_{\x} } 
\right| _{\rho_{\y}, w_{\x\y}^2}\, , \label{eq:defmu} \\
w^{\x\y}_{i}     & \equiv & v_{i}^{\x} -  v_{i}^{\y}  \, .
\end{eqnarray}
The energy functional $\mathcal{E}=\mathcal{E}(n_\n,n_\p,w_{\n\p}^2)$
describes the equation of state~(EoS) of the system. Finally, the
non-dissipative entrainment between the two fluids is governed by the
parameter $\varepsilon_{\x}$, which follows from the definition:
\begin{eqnarray}
&& \varepsilon_{\x} \equiv 2 \rho _{\x}\left.  \frac{\partial
\mathcal{E}}{\partial w^{2}_{\n\p} } \right|_{\rho_\x,\rho_\y} \, . \label{eq:vareps}
\end{eqnarray}

The equations that describe rapidly and uniformly rotating background
models can be derived by integrating the Euler-type
equations~(\ref{eq:Euler}) and the Poisson
equation~(\ref{eq:Poisson})~\citep{2002A&A...381..178P,2004MNRAS.347..575Y}. This
leads to
\begin{equation}
 \tilde \mu _\x + \Phi - \frac{r^2}{2} \sin \theta ^2 \, \Omega_\x^2  =  C_\x \, , \label{eq:bgmu_x} 
\end{equation}
where $\Omega_\x$ and $C_\x$ are, respectively, the angular velocities
and the integration constants for the neutron and proton fluids.
In this work, we focus on corotating background models,
$\Omega_{\n}=\Omega_{\p}=\Omega$, where the two fluids are in
$\beta$-equilibrium and have a common surface.  The hydrostatic
equilibrium equation (\ref{eq:bgmu_x}) then
become
\begin{equation}
\tilde \mu + \Phi - \frac{r^2}{2} \sin \theta ^2 \, \Omega^2  =  C \, , \label{eq:bg1}
\end{equation}
where $\tilde{\mu}_{\p} = \tilde{\mu}_{\n} \equiv \tilde{\mu}$ is the
background chemical potential and $C\equiv C_\n = C_\p$. It is worth
noticing that equations~(\ref{eq:Poisson}) and~(\ref{eq:bg1}) are
formally equivalent to the single fluid problem provided one replaces
the chemical potential with the
enthalpy~\citep{2004MNRAS.347..575Y}. Given an equation of state,
equations~(\ref{eq:Poisson}) and~(\ref{eq:bg1}) can be numerically
solved using the self-consistent field method
of~\citet{1986ApJS...61..479H}. The surface of the star corresponds to
the zero chemical potential surface, $\tilde {\mu} \left(
r\left(\theta\right) , \theta \right) =
0$~\citep{2004MNRAS.347..575Y}.

%%%%%%%%%%%%%%%%%%%%%%%%%%%%% Sec %%%%%%%%%%%%%%%%%
\section{Perturbation Equations}
%%%%%%%%%%%%%%%%%%%%%%%%%%%%%%%%%%%%%%%%%%%%%%%%%%%%

It is straighforward to write down the system of partial differential
equations that governs the Eulerian perturbations $\delta \rho_{\x} ,
\delta \mtb{v} _{\x}, \delta \tilde{\mu}_{\x}$ and $\delta
\Phi$. However, instead of working with these variables, we
define~\citep*[see][for a detailed discussion]{2008arXiv0812.3023A}
new  variables which are more closely related to the natural
degrees of freedom of the problem. In the rotating frame, the
co-moving (ordinary) motion is described by the mass flux $\mtb{f}$
(not to be confused with the mutual force $\mtb{f}^{\x}$), the total
mass density $\delta \rho$ and the pressure $\delta P$, defined by
\begin{eqnarray}
\mtb{f} & = & \rho_\p \mtb{v}_\p + \rho_\n \mtb{v}_\n \, , \label{eq:def-f} \\
\delta \rho & = & \delta \rho _{\n} + \delta \rho_{\p} \, , \\
\nabla \delta P & = & \rho_\p \nabla \delta \tilde \mu _\p 
+ \rho_\n \nabla \delta \tilde \mu _\n \, . \label{eq:def-dP}
\end{eqnarray}
Meanwhile, the counter-moving (superfluid) motion is described by a
vector field $\mtb{D}$ that is proportional to the relative velocity
between the two fluids, the scalar perturbation $\delta \beta$ that
measures the deviation from $\beta$-equilibrium and the quantity
$\delta \chi_\p$, which is related to the perturbed proton fraction.
These variables are defined by
\begin{eqnarray}
\mtb{D} & = & x_\p \left( 1 - x_\p \right) \rho \left( \mtb{v}_\p - \mtb{v}_\n\right) \, , \label{eq:def-D}  \\
\delta \beta & = &  \delta \tilde \mu _\p - \delta \tilde \mu _\n \, , \label{eq:def-dmu} \\
\delta \chi_\p & = & \rho \, \delta x_\p \, , \label{eq:def-dchip}
\end{eqnarray}
where $x_\p = \rho_\p / \rho$ is the proton fraction.

In order to simplify the evolutions, we neglect the perturbations of
the gravitational potential $\delta \Phi = 0$. That is, we adopt the
Cowling approximation. This approximation should be quite accurate for
inertial modes.  For low order acoustic modes, like the f-mode, it is
not so accurate but the results are still qualitatively correct.  For
our present purposes, this should be sufficent. Although, it would not
be too difficult to solve also for the perturbed gravitational
potential it is computationally costly to add the solution of an
elliptic equation to our evolutions. Hence, we decided not to solve the full
problem in this first exploratory study.

In the frame of the rotating background the final perturbation
equations can be written
\begin{eqnarray}
\partial_t \mtb{f}    & = & - \nabla \delta P - 2 \mtb{\Omega} \times \mtb{f}
                        + \frac{\nabla P}{\rho} \,  \delta \rho  \, ,                       \label{eq:dfdt} \\
 \partial_t \mtb{D}    & = &  - \frac{x_\p \left(1-x_\p \right) \rho  \nabla \delta \beta }{ 1 - \bar \varepsilon }
                            - \frac{ 2 \mtb{\Omega} \times \mtb{D} }{ 1 - \bar \varepsilon }   \, , \label{eq:dDdt} \\
\partial_t\delta \rho & = & -  \nabla \cdot  \mtb{f}   \, ,        \label{eq:drhodt} \\
\partial_t\delta \chi _\p & = & - \nabla \cdot \mtb{D} - \mtb{f} \cdot \nabla  x_\p  \, ,  \label{eq:dchipdt}
\end{eqnarray}
where we have defined $\bar \varepsilon = \varepsilon_\n / x_\p$.

The time evolution of the non-axisymmetric perturbation equations is a
three-dimensional problem in space. However, linear perturbations on
an axisymmetric background can be expanded in terms of a set of basis
functions $\left( \cos m \phi \, ,\sin m \phi \right)$, where $m$ is
the azimuthal harmonic index~\citep{1980MNRAS.190...43P}. The mass
density perturbations as well as the other perturbation quantities
take the following form~\citep{2002MNRAS.334..933J,
2008arXiv0807.3457P}
\begin{equation}
\delta \rho \left( t,r,\theta,\phi \right) = \sum_{m=0}^{m=\infty}
               \left[ \delta \rho_{ m}^{+} \left( t,r,\theta\right)
               \cos m \phi + \delta \rho_{m}^{-} \left(
               t,r,\theta\right) \sin m \phi \right] \, .
               \label{eq:drhoexp}
\end{equation}
The perturbation equations now decouple with respect to $m$ and the
problem becomes two-dimensional. Therefore for any $m$, the
non-axisymmetric oscillations of a superfluid neutron star requires
the solution of a system of eighteen partial differential equations
for the twenty variables $\left( \mtb{f}^{\pm}, \delta \rho ^{\pm},
\delta P ^{\pm}, \mtb{D}^{\pm}, \delta \chi_\p^{\pm}, \delta \beta
^{\pm} \right)$. To fully specify the problem, the set of
equations~(\ref{eq:dfdt})-(\ref{eq:dchipdt}) must be complemented by
two relations that depend on the EoS (see Section~\ref{sec:EoS}).

%%%%%%%%%%%%%%%%%%%%%%%%%%%%%%% SEC. %%%%%%%%%%%%%%%%%%%%%%%%%%%%%%%%%%%%%%%%%%%
\subsection{Boundary Conditions} \label{sec:BC}
%%%%%%%%%%%%%%%%%%%%%%%%%%%%%%%%%%%%%%%%%%%%%%%%%%%%%%%%%%%%%%%%%%%%%%%%%%%%%%%%

In order to evolve the perturbation equations we must also specify
boundary conditions.  For non-axisymmetric oscillations with $m \geq
2$, equations~(\ref{eq:dfdt})--(\ref{eq:dchipdt}) are regular at the
origin, $r=0$, when the following conditions are satisfied:
\begin{equation}
\delta P = \delta \chi_\p =  \delta \beta = \delta
\rho = 0 \, , \qquad \textrm{and}  \qquad \mtb{f} = \mtb{D} = \mtb{0} \, .
\end{equation}

For the boundary condition at the stellar surface, it is worth
remembering that the unperturbed configuration is such that the two
fluids have a common surface~(see Section~\ref{sec:Intro}). At the
perturbed level, we require that the Lagrangian perturbation of the
 chemical potentials vanishes at the surface, i.e.,
\begin{equation}
\Delta_\x \tilde{\mu}_{\x} = \delta \tilde{\mu}_{\x}
+ \mtb{\xi}_{\x}  \cdot \nabla \tilde{\mu}_{\x} = 0  \, , \label{eq:Dmu-bc}
\end{equation}
where the Lagrangian variations $\Delta_\x$ are associated with the
fluid displacements $\mtb{\xi}_{\x}$~\citep*{2004MNRAS.355..918A}.
Equations~(\ref{eq:Dmu-bc}) can be expressed in terms of $\delta P$
and $\delta \beta$ by using the definitions~(\ref{eq:def-dP})
and~(\ref{eq:def-dmu}). This leads to
\begin{eqnarray}
\Delta P & = & \delta P + \left[ x_{\p} \, \mtb{\xi}_{\p}
+ \left( 1 - x_{\p} \right) \mtb{\xi}_{\n} \right]   \cdot \nabla P = 0  \, , \label{eq:DP-bc} \\
\Delta \beta  & = & \delta \beta  + \left( \mtb{\xi}_{\p}
- \mtb{\xi}_{\n } \right) \cdot \frac{\nabla P}{\rho} = 0  \, , \label{eq:Dbeta-bc}
\end{eqnarray}
where we have used the fact that the background model is in
$\beta$-equilibrium condition, i.e., $\tilde{\mu}_{\n} =
\tilde{\mu}_{\p}$.
As in the single fluid case we can derive a simpler condition for the
pressure perturbation.  From the Euler equation for the stationary
background (noting that (\ref{eq:def-dP}) holds also at the
unperturbed level)
\begin{equation}
\nabla P = - \rho \mtb{\Omega} \times \left( \mtb{\Omega} \times \mtb{r} \right)
           - \rho \nabla \Phi   \, ,  \label{eq:Euler-bck}
\end{equation}
and the fact that the EoS used in this work are such that the total
mass density vanishes at the surface, it follows that
equation~(\ref{eq:DP-bc}) is equivalent to~\citep{Tassoul:1978}:
\begin{equation}
\delta P = 0 \, .
\end{equation}
This condition is numerically convenient, and ensures that all
variables remain regular at the surface.

The reflection symmetry with respect to the equator divides the
perturbation variables into two
sets~\citep{2008arXiv0807.3457P}. Perturbations of the Type~I parity
class have $\left( f_r, f_{\phi}, \delta \rho, \delta P , D_r,
D_{\phi} , \delta \chi_{\p} , \delta \beta \right)$ even and $\left(
f_{\theta} , D_{\theta}\right)$ odd. The opposite is true for
perturbations of Type II.

%%%%%%%%%%%%%%%%%%%%%%%%%%%%%%% SEC. %%%%%%%%%%%%%%%%%%%%%%%%%%%%%%%%%%%%%%%%%%%
\section{Equation of State} \label{sec:EoS}
%%%%%%%%%%%%%%%%%%%%%%%%%%%%%%%%%%%%%%%%%%%%%%%%%%%%%%%%%%%%%%%%%%%%%%%%%%%%%%%%

To complete the formulation of the superfluid oscillation problem we
need to supply a suitable multi-parameter equation of state. In a
truly realistic model, the equation of state should be obtained from a
microscopic (quantum) analysis. It will completely specify not only
the relation between pressure and density, but also the composition
and the detailed superfluid energy gaps for neutrons and
protons. Moreover, it has to also provide the entrainment parameters.
We do not yet have such a model, although recent developments in this
direction are promising~\citep{2008MNRAS.388..737C}. Given this, and
the fact that our main aim is to understand the oscillations of a
rotating superfluid neutron star at the qualitative level, we will opt
to work with two simple analytic model equations of state. These
models, described in Sections~\ref{sec:BGEoS_1} and
~\ref{sec:BGEoS_2} below, are natural generalisations of the single fluid
polytropes.  The analytical models are particularly useful since they
allow us to tune key parameters like entrainment and symmetry energy
more or less freely. As we will see, we can also vary the coupling
between the co- and counter-moving fluid degrees of freedom.

Quite generally, the required energy functional can be expressed as a
function of the two fluid mass densities and the relative velocity:
\begin{equation}
\mathcal{E} = \mathcal{E} \left( \rho_\n, \rho_\p , w_{\n \p}^2 \right) \, , \label{eq:EoS}
\end{equation}
where the dependence on $\mtb{w}_{\n\p}$ ensures Galilean
invariance. For a small relative velocity between the two fluids,
equation~(\ref{eq:EoS}) can be written
\begin{equation}
\mathcal{E} = \mathcal{E}_{0} \left(\rho_\n, \rho_\p \right)
+ \alpha_0 \left( \rho_\n, \rho_\p \right) w_{\n \p}^2 + \mathcal{O}\left(w_{\n \p}^4\right) \, , \label{eq:EoSbulk}
\end{equation}
in which case the bulk EoS $\mathcal{E}_{0}$ and the entrainment
$\alpha_0$ can be independently specified at
$\mtb{w}_{\n\p}=\mtb{0}$. From equation~(\ref{eq:vareps}) follows that
the entrainment parameter $\varepsilon _{\x}$ is related to the
function $\alpha_0$ by
\begin{equation}
\rho_\x  \varepsilon_\x = 2 \alpha_0 \, .  \label{eq:alp}
\end{equation}

Having specified the EoS, we must determine two equations that close
the system~(\ref{eq:dfdt})--(\ref{eq:dchipdt}).  One possible choice
is to determine the pressure perturbation and the quantity $\delta
\beta $ from the total density and the proton fraction
\begin{equation}
\delta P = \delta P \left( \delta \rho , \delta x_\p \right) \, ,
\qquad \qquad \delta \beta = \delta \beta \left( \delta \rho , \delta
x_\p \right) \, .  \label{eq:EoSpr}
\end{equation}
The required relations are obtained by first expressing these
quantities in terms of the  chemical potential perturbations
$\delta \tilde \mu _{\x}$, i.e. using the thermodynamic definitions
\begin{eqnarray}
\delta P     & = & \rho_\p \delta \tilde \mu_\p + \rho_\n \delta \tilde \mu_\n \, , \label{eq:dPdmu}\\
\delta \beta & = & \delta \tilde \mu_\p - \delta \tilde \mu_\n \, .  \label{eq:dbetadmu}
\end{eqnarray}
For corotating background models~($\mtb{w}_{\n\p}=\mtb{0}$), the
perturbation of the chemical potential $\tilde \mu_\x = \tilde \mu_\x
\left( \rho_{\p} , \rho_{\n} \right)$ can be expressed as
\begin{equation}
\delta \tilde \mu _{\x} = \left. \frac{ \partial \tilde \mu_{\x} }{ \partial \rho_{\p} } \right|_{\rho_\n} \delta \rho_{\p}
                        + \left. \frac{ \partial \tilde \mu_{\x} }{ \partial \rho_{\n} } \right|_{\rho_\p} \delta \rho_{\n} \, ,
\label{eq:tildemu}
\end{equation}
where the mass densities of the two fluid components are defined in
terms of the total mass density and proton fraction by
\begin{eqnarray}
\delta \rho_\p & = & x_\p \delta \rho + \rho \, \delta x_\p \, , \label{eq:drhop} \\
\delta \rho_\n & = & \left( 1 - x_\p \right) \delta \rho - \rho \, \delta x_\p \, . \label{eq:drhon}
\end{eqnarray}
By introducing equations~(\ref{eq:tildemu})--(\ref{eq:drhon}) into
equations~(\ref{eq:dPdmu})--(\ref{eq:dbetadmu}), we obtain:
\begin{eqnarray}
\delta P     & = &    \left[ \left( \left(1+2\sigma \right) x_\p ^2 - 2 \left( 1 + \sigma \right) x_\p + 1 \right)  A_{\n\n}
               + x_\p^2  A_{\p\p} \right]  \rho \, \delta \rho
               + \left[ \left( \left(1+2\sigma \right) x_\p - 1 - \sigma \right) A_{\n\n}
               +  x_\p  A_{\p\p} \right]  \rho \, \delta \chi_\p \, , \label{eq:dPeos}  \\
\delta \beta  & = & \left[ \left( \left(1+2\sigma \right) x_\p - 1 - \sigma \right) A_{\n\n}
               +     x_\p A_{\p\p} \right]  \delta \rho
               +  \left[  \left( 1+2 \sigma \right) A_{\n\n} + A_{\p\p} \right] \delta \chi_\p \, ,  \label{eq:dbeos}
\end{eqnarray}
where the matrix $A_{\x \y}$ is defined by
\begin{equation}
A_{\x \y} \equiv \frac{ \partial \tilde \mu_{\x} }{ \partial \rho_{\y}
} = \frac{ \partial ^2 \mathcal{E} }{ \partial \rho_{\y} \partial
\rho_{\x} } \, , \label{eq:Axy}
\end{equation}
and $\sigma$ corresponds to the so-called ``symmetry
energy''~\citep*{2002A&A...381..178P, 1988PhRvL..61.2518P}. That is,
we have
\begin{equation}
\sigma \equiv - \frac{A_{\n\p}}{A_{\n\n}} \, .
\end{equation}

%%%%%%%%%%%%%%%%%%%%%%%%%%%%%%%%%%% Sec %%%%%%%%%%%%%%%%%%%%%%%%%%%%%%%%%%%%%%%
\section{Stellar Models} \label{sec:St-We}
%%%%%%%%%%%%%%%%%%%%%%%%%%%%%%%%%%%%%%%%%%%%%%%%%%%%%%%%%%%%%%%%%%%%%%%%%%%%%%

Background stellar models such that the two fluids are co-rotating can
be constructed by solving the hydrostatic equilibrium
equations~(\ref{eq:Poisson}) and~(\ref{eq:bg1}) for a given bulk EoS
$\mathcal{E}_{0}$, cf. ~(\ref{eq:EoSbulk}).  Since
equations~(\ref{eq:Poisson}) and~(\ref{eq:bg1}) are formally
equivalent to the equilibrium equations for a single fluid polytrope,
we can straightforwardly use the method of~\citet{1986ApJS...61..479H}
to determine such background models~(see Section~\ref{sec:pert-eqs}).

For a co-rotating background, entrainment does not not affect the
equilibrium configuration. Hence, it can be chosen independently from
the bulk EoS (see equations~(\ref{eq:EoSbulk}) and~(\ref{eq:alp})). In
fact, the entrainment parameter appears only in the perturbation
equation~(\ref{eq:dDdt}) through the combination
\begin{equation}
\bar \varepsilon = \frac{\varepsilon_\n}{ x_\p} = \varepsilon_\p + \varepsilon_\n \, .
\end{equation}
In the last step we have used the relation~(\ref{eq:alp}),
i.e. $\rho_\n \varepsilon_\n = \rho_\p \varepsilon_\p$.
Nuclear physics calculations limit the value of the entrainment in the
neutron star core to~$0.2 \leq \varepsilon_\p \leq 0.8$, see
\citet{2008MNRAS.388..737C} for a recent analysis.  However, values outside
this range are possible, especially for the crust superfluid. In fact,
the parameter $\varepsilon_\p$ is expected to have negative values in
the crust region~\citep{2006NuPhA.773..263C}. Since we are interested
in exploring the effect that the different parameters have on the
neutron star oscillation modes, we will consider the range $- 0.7 \leq
\bar \varepsilon \leq 0.7$.

%~~~~~~~~~~~~~~~~~~~~~~~~~~~~~~~~~~~~~~~~~~~~~~~~~~~~~~~~~~~~~~~~~~~~~~~~~~~~~
\subsection{ Model A} \label{sec:BGEoS_1}
%~~~~~~~~~~~~~~~~~~~~~~~~~~~~~~~~~~~~~~~~~~~~~~~~~~~~~~~~~~~~~~~~~~~~~~~~~~~~~
%------------------------------TAB. 1------------------------------------------%
\begin{table}
\begin{center}
\caption{\label{tab:back-models} This table provides the main
  parameters for the rotating models A and B~(see
  Sections~\ref{sec:BGEoS_1} and~\ref{sec:BGEoS_2} for the detailed
  EoS).  The first column labels each model. In the second and third
  columns we show, respectively, the ratio of polar to equatorial axes
  and the angular velocity of the star. In the fourth column, the
  rotation rate is compared to the Kepler velocity $\Omega_K$ that
  represents the mass shedding limit. The ratio between the rotational
  kinetic energy and gravitational potential energy $T/|W|$ and the
  stellar mass are given in the fifth and sixth columns,
  respectively. All quantities are given in dimensionless units, where
  $G$ is the gravitational constant, $\rho_c$ represents the central
  mass density and $R_{eq}$ is the equatorial radius.}
\begin{tabular}{ c c  c c c c  }
\hline
 Model &  $ R_p / R_{eq} $  &  $ \Omega / \sqrt{G\rho_c}$ & $\Omega / \Omega_K$  & $ T/|W| \times 10^{2}$
  & $ M / (\rho_c R_{eq}^3)$  \\
\hline
A0  &  1.000            &         0.000     &     0.000      &    0.000   &    1.273   \\
A1  &  0.996            &         0.084     &     0.116      &    0.096   &    1.270   \\
A2  &  0.983            &         0.167     &     0.230      &    0.385   &    1.248   \\
A3  &  0.950            &         0.287     &     0.396      &    1.169   &    1.197   \\
A4  &  0.900            &         0.403     &     0.556      &    2.385   &    1.118   \\
A5  &  0.850            &         0.488     &     0.673      &    3.639   &    1.038   \\
A6  &  0.800            &         0.556     &     0.767      &    4.933   &    0.956   \\
A7  &  0.750            &         0.612     &     0.844      &    6.252   &    0.869   \\
A8  &  0.700            &         0.658     &     0.907      &    7.568   &    0.779   \\
A9  &  0.650            &         0.693     &     0.956      &    8.822   &    0.684   \\
\hspace{0.5mm} A10 &  0.600 &     0.717     &     0.989      &    9.865   &    0.579   \\
\hspace{0.5mm} A11 &  0.558 &     0.725     &     0.999      &   \hspace{-0.18cm}10.277   &    0.480   \\
\hline
B0 &  1.000            &         0.000      &     0.000      &    0.000   &    1.833   \\
B1 &  0.996            &         0.094      &     0.107      &    0.105   &    1.825   \\
B2 &  0.983            &         0.187      &     0.213      &    0.419   &    1.799   \\
B3 &  0.950            &         0.323      &     0.368      &    1.275   &    1.733   \\
B4 &  0.900            &         0.453      &     0.516      &    2.608   &    1.632   \\
B5 &  0.850            &         0.551      &     0.628      &    4.002   &    1.529   \\
B6 &  0.800            &         0.629      &     0.717      &    5.459   &    1.424   \\
B7 &  0.750            &         0.695      &     0.792      &    6.981   &    1.317   \\
B8 &  0.700            &         0.751      &     0.856      &    8.566   &    1.207   \\
B9 &  0.650            &         0.798      &     0.909      &   \hspace{-0.18cm}11.020   &    1.093   \\
\hspace{0.5mm} B10  &   0.600 &   0.835     &     0.952      &   \hspace{-0.18cm}11.862   &    0.973   \\
\hspace{0.5mm} B11  &   0.563 &   0.857     &     0.977      &   \hspace{-0.18cm}13.074   &    0.875   \\
\hspace{0.5mm} B12  &  0.496 &   0.877      &     0.999      &   \hspace{-0.18cm}14.613   &    0.669   \\
\hline
\end{tabular}
\end{center}
\end{table}
%------------------------------------------------------------------------------%

As our first model equation of state we consider
~\citep[see,][]{2002A&A...381..178P, 2004MNRAS.347..575Y}
\begin{equation}
\mathcal{E}_0 = \frac{1}{2} \sum_{\x,\y} A_{\x\y} \rho_\x \rho_\y \, .
\label{eq:EoS1}
\end{equation}
We will refer to this as model A. Despite its obvious simplicity, this
model allows us to investigate the effect of the symmetry energy on
the oscillation spectrum. This is apparent from the matrix
coefficients $A_{\x\y}$ which have the following form:
\begin{eqnarray}
A_{\n\n} &=& \frac{2 K}{1-\left( 1 + \sigma \right) x_\p} \label{eq:AxyPCA1}\ , \\
A_{\p\p} &=& \frac{2 K  \left[ 1 + \sigma - \left( 1 + 2 \sigma \right) x_\p \right] }
{x_\p \left[  1-\left( 1 + \sigma \right) x_\p \right] } \label{eq:AxyPCA2}
\, ,\\
\qquad A_{\n \p} &=& - \sigma A_{\n \n} \, , \label{eq:AxyPCA3}
\end{eqnarray}
where~$\sigma$ and $K$ are
constants. Equations~(\ref{eq:AxyPCA1})--(\ref{eq:AxyPCA3}) are
equivalent to the set used by~\citet{ 2004MNRAS.347..575Y} with $2K =
1/k$.  From the $\beta$-equilibrium condition for the stellar
background, $\tilde{\mu}_\n = \tilde{\mu}_\p$, and the
definition~(\ref{eq:defmu}) we can derive the proton fraction for
these models
\begin{equation}
x_\p = \frac{ \left( 1 + \sigma \right) A_{\n\n}}{ A_{\p\p} + \left( 1
+ 2 \sigma \right) A_{\n\n}} \, .
\end{equation}
If we take the coefficients~$A_{\x\y}$ to be constant, model A leads
to a family of non-stratified stars.  The perturbation variables are
related by equations~(\ref{eq:dPeos})--(\ref{eq:dbeos}) which in this
case become
\begin{equation}
\delta P     = 2 K \rho \, \delta \rho \, , \qquad \qquad
\delta \beta = \frac{ 2 K \left( 1+\sigma \right) }
                { x_\p \left[ 1 - \left( 1 + \sigma \right) x_\p \right] }  \label{eq:deos1}
\delta \chi_\p \, .
\end{equation}
Therefore, the models are specified by the three parameters $K,
\sigma$ and $x_\p$.  For this class of models, the degrees of freedom
that describe the co- and counter-moving fluid motion are decoupled.
The variables $\mtb{f}, \delta \rho, \delta P$ evolve independently
from $\mtb{D}, \delta \chi_\p , \delta \beta$. This is a useful
simplification that helps the interpretation of the oscillation
spectrum.

For a background star in $\beta$-equilibrium, the  chemical
potential is related to the total mass density by
\begin{equation}
\tilde \mu = 2 K \rho \, ,
\end{equation}
which means that the pressure is that of the usual $N=1$~polytrope:
\begin{equation}
P = K \rho ^2 \, . \label{eq:N1EoS}
\end{equation}

We have constructed a family of rotating stars for this equation of
state.  In dimensionless units, the constant $K$ is determined
automatically by specifying the polytropic index and the axis ratio
between the polar and equatorial radius $R_p /
R_{eq}$~\citep{2002MNRAS.334..933J, 2008arXiv0807.3457P}. The
properties of our rotating models are given in Table\footnote{ In
compiling Table 1, we noticed an error in the data reported
in~\cite{2008arXiv0807.3457P}.  The data given in the fourth column of
Table 1 of that paper is incorrect. For the sequence of stellar model
A, the correct value of the break-up angular velocity is $\Omega_K /
\sqrt{ G \rho_c}= 0.7252$ and the present table gives the correct
values for the related quantity $\Omega / \Omega_K$. We also note a
typo in the label of the fifth column in the previous paper, where
$T/|W|\times 10^{-2}$ should be replaced by $T/|W|\times
10^{2}$.}~\ref{tab:back-models}, where we also label each member of
the sequence. The non-rotating model is refered to as A0 and the
fastest rotating model is A11. Being non-stratified stars, the values
of both the proton fraction $x_\p$ and the symmetry energy term
$\sigma$ affect only the counter-moving motion of the perturbed
fluid. In fact, they appear only in the perturbation
equations~(\ref{eq:dDdt}) and~(\ref{eq:dchipdt}), and in
equation~(\ref{eq:deos1}) for the $\delta \beta$ perturbation. In our
numerical simulations, we focus mainly on stars with $x_\p=0.1$ and
$-1 \le \sigma \le 1$. The range of the symmetry energy is constrained
by the requirement that $ A_{\x\y}$ should be invertible~\citep[see
e.g.,][]{2002A&A...381..178P} as well as realistic calculations for
neutron star EoS~\citep{2007PhR...442..109L}.  For simplicity, we will
assume that the symmetry energy is constant throughout the star.

%~~~~~~~~~~~~~~~~~~~~~~~~~~~~~~~~~~~~~~~~~~~~~~~~~~~~~~~~~~~~~~~~~~~~~~~~~~~~~
\subsection{ Model B} \label{sec:BGEoS_2}
%~~~~~~~~~~~~~~~~~~~~~~~~~~~~~~~~~~~~~~~~~~~~~~~~~~~~~~~~~~~~~~~~~~~~~~~~~~~~~

Our second model equation of state is constructed in such a way that
we can explore the relevance of the chemical coupling between the two
fluid degrees of freedom that arises due to composition variation. We
consider the simple analytical EoS~\citep{2002A&A...393..949P,
2002PhRvD..66j4002A}:
\begin{equation}
\mathcal{E}_{0} = k_{\n} \, \rho_\n^{\gamma_n} + k_{\p} \,
\rho_\p^{\gamma_\p} \, , \label{eq:EosPR}
\end{equation}
where $k_\x$ are constant coefficients. For this EoS, the symmetry
energy term vanishes, as $A_{\n\p} = 0$. However, the polytropic
indices $N_\x = \left(\gamma_\x -1 \right) ^{-1}$ of the neutron and
proton fluids can be different, i.e. $N_\n \neq N_\p$. This enables us
to construct stratified configurations. To see this, consider the
relation between the chemical potential and the mass density
\begin{equation}
\rho_\x = \left( \frac{\tilde \mu _\x }{k_\x \gamma_\x } \right) ^{N_\x} \, ,
\end{equation}
which for a model in $\beta$-equilibrium leads to the following
profile for the proton fraction:
\begin{equation}
x_\p = \left[ 1 + \frac{ \left( \gamma_\p k_\p \right) ^{N_\p} }{\left( \gamma_\n k_\n \right) ^{N_\n}}
\, \tilde \mu ^{N_\n-N_\p} \right]^{-1} \, .  \label{eq:xp}
\end{equation}
For a vanishing symmetry energy,
equations~(\ref{eq:dPeos})--(\ref{eq:dbeos}) become
\begin{eqnarray}
\delta P  & = & \left[ \left(x_\p ^2 - 2 x_\p + 1 \right) A_{\n\n} + x_\p^2  A_{\p\p} \right]
             \rho \, \delta \rho
             + \left[  \left(x_\p-1 \right) A_{\n\n} + x_\p  A_{\p\p} \right] \rho
             \, \delta \chi_\p \, , \label{eq:dPeosPR}  \\
\delta \beta  & = & \left[ \left( x_\p - 1 \right) A_{\n\n} + x_\p A_{\p\p} \right] \delta \rho
               + \left(  A_{\p\p} + A_{\n\n} \right) \delta \chi_\p \, .
             \label{eq:dbeosPR}
\end{eqnarray}
For model B the $A_{\x\y}$ coefficients are explicitly given by
\begin{equation}
A_{\x \x} = k_\x \gamma_\x \left( \gamma_\x - 1 \right) \rho_\x ^{\gamma_\x -2} \, . \label{eq:AxyPR}
\end{equation}

We have constructed two different non-rotating models for this
analytic EoS.  These two models are labeled $\rm B^{NS}$ and B0, and
correspond to (after a transformation of units) models I and II
of~\citet{2002A&A...393..949P}. The parameters for models~$\rm B^{NS}$
and B0 are given in Table~\ref{tab:PR-models}.  Comparing model I and
II of~\citet{2002A&A...393..949P} to our numerical models, we find an
agreement to better than one percent for the dimensionless stellar
mass $M / \left( \rho_c R_{eq}^3\right)$.
%
%------------------------------TAB. 2------------------------------------------%
\begin{table}
\begin{center}
\caption{\label{tab:PR-models} This table provides the parameters for
  two non-rotating stellar models for the EoS~(\ref{eq:EosPR}).  We
  refer to these models as $\rm B^{NS}$ and B0. They correspond to 
  models I and II of~\citet{2002A&A...393..949P}, respectively. The
  units of the coefficients $k _\x$ are $ G R_{eq}^2
  \rho_c^{2-\gamma_\x}$, where $G$ is the gravitational constant and
  $R_{eq}$ is the equatorial radius. The proton fraction at the star's
  centre is $x_\p\left(0\right)$, while the central mass density is
  $\rho_c$. }
\begin{tabular}{c c c c c c c  }
\hline
 Model &  $ \gamma_{\n} $  &  $ \gamma_{\p} $  & $k_\n  $  & $k_\p$ &
  $ x_\p\left(0\right) $   & $ M / \left( \rho_c R_{eq}^3\right)  $   \\
\hline
 $\rm B^{NS}$   &  2.0      &  2.0  & 0.705  & 6.347  & 0.1    & 1.273  \\
 \hspace{-3mm} $\rm B0$     &  2.5  &  2.1   & 0.693 & 8.871  & 0.1    & 1.833  \\
\hline
\end{tabular}
\end{center}
\end{table}
%------------------------------------------------------------------------------%

The non-rotating $\rm B^{NS}$ model represents a non-stratified star,
$\gamma_p = \gamma_n$, with a constant proton fraction given by
\begin{equation}
x_\p = \frac{k_\n}{ k_\n + k_\p } = 0.1\, .
\end{equation}
This means that, the co- and counter-moving perturbations are
decoupled and equations~(\ref{eq:dPeosPR})--(\ref{eq:dbeosPR}) become:
\begin{equation}
\delta P = 2 \frac{k_\n k_\p}{k_\n + k_\p} \rho \, \delta \rho \, ,
\qquad \qquad \delta \beta = 2 \left( k_\n + k_\p \right) \delta
\chi_\p \, . \label{eq:deos2I}
\end{equation}
We use this model to test our evolutions against the frequency domain results
of~\citet{2002A&A...393..949P}.

In addition we consider a rotating sequence, which extends B0 up to
the fastest rotating model B12.  All our rotating B models correspond
to the same~$\gamma_x$ and~$k_\x$ as the B0 model. Therefore, any
rotating B model is stratified with $x_\p(0) \neq 0$ at the centre and
zero proton fraction at the star's surface. Due to the effect of
rotation on the central density and chemical potential, the central
proton fraction is $x_\p(0) = 0.1$ for the non-rotating model B0 and
becomes $x_\p(0) \simeq 0.081$ for model B12, which is near the mass
shedding limit. These stratified models are physically interesting, as
the relations~(\ref{eq:dPeosPR})--(\ref{eq:dbeosPR}) and the
perturbation equations couple the co- and counter-moving degrees of
freedom and we can study the effect of this coupling on the spectrum
of stellar oscillations.  The main properties of the B models are
given in Table~\ref{tab:back-models}.

%------------------------------FIG. 1------------------------------------------%
\begin{figure}
\begin{center}
\includegraphics[height=75mm]{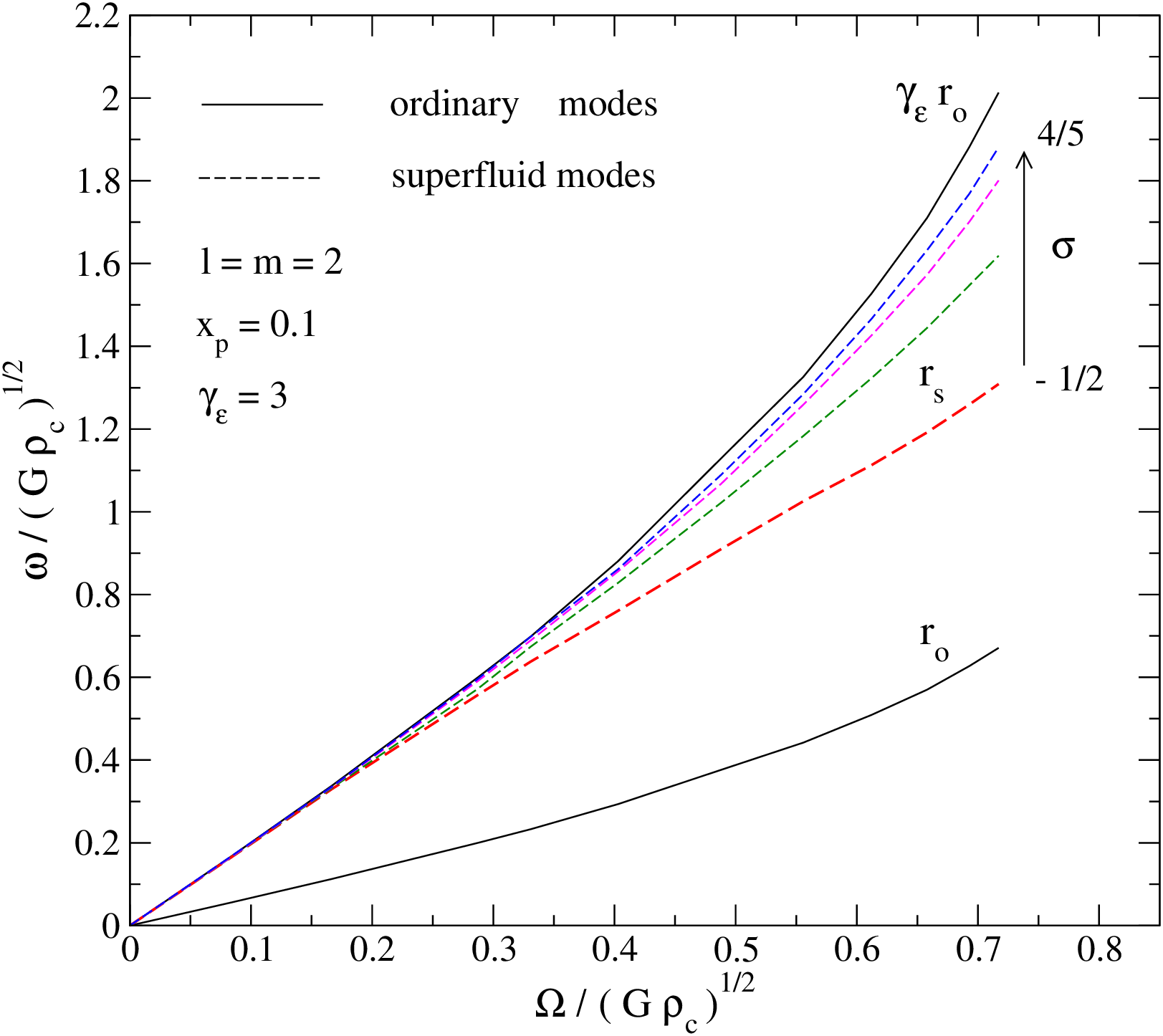}
\includegraphics[height=75mm]{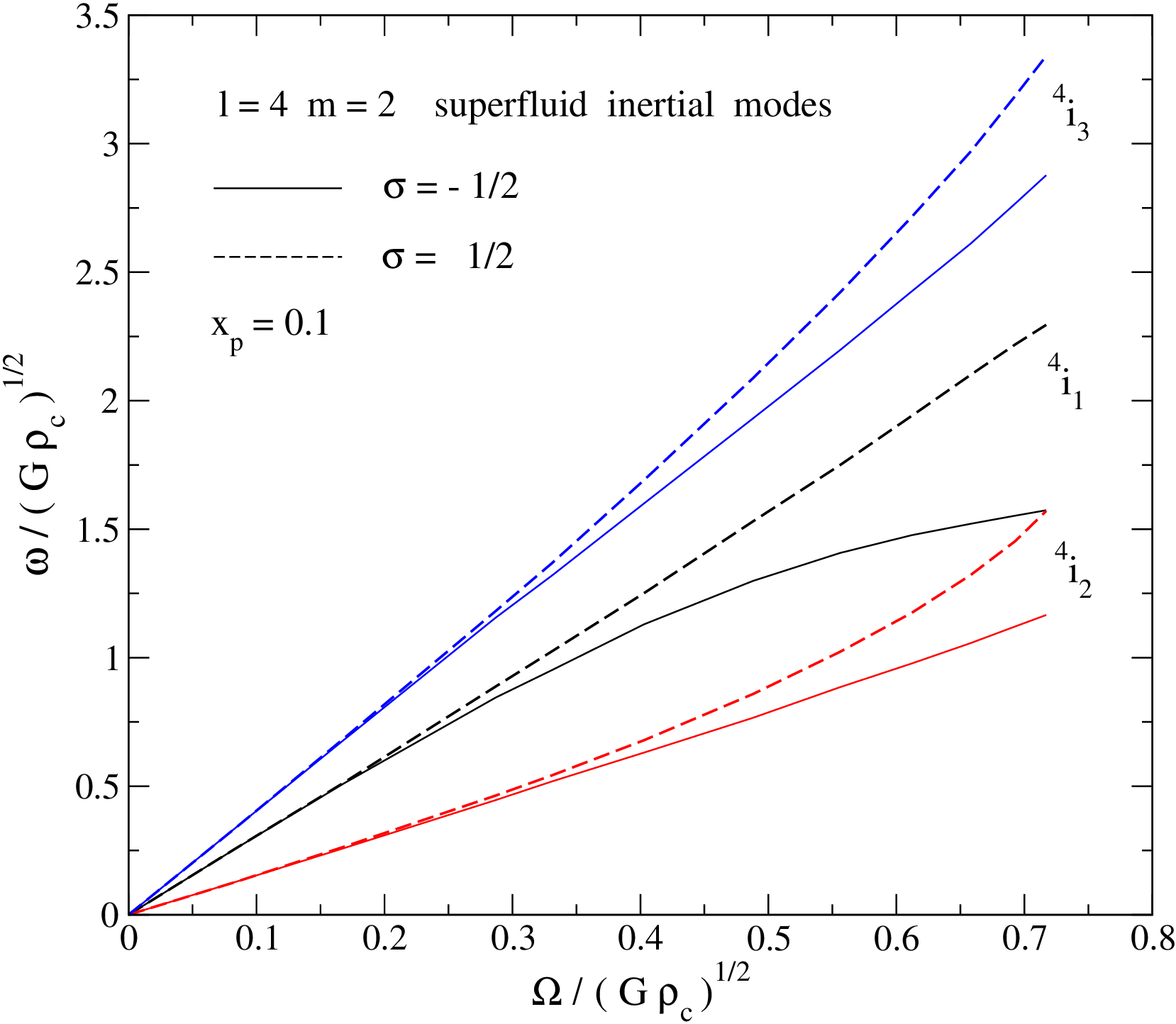}
\caption{This figure shows the frequencies of the axial-led inertial
  modes and their dependence on the star's spin and the symmetry
  energy~$\sigma$ for the sequence of rotating models A with proton
  fraction~$x_\p = 0.1$ and entrainment parameter~$\bar \varepsilon =
  2/3$.  The mode-frequencies and the stellar angular velocity are
  given in units of $\sqrt{ G \rho_c}$, where $G$ is the gravitational
  constant and $\rho_c$ the central mass density. The mode-frequencies
  are determined in the frame of the rotating star.  In the left
  panel, we show the ordinary $\rm r_o$-mode and the superfluid $\rm
  r_\s$-modes for different values of $\sigma$ in the range $-1/2 \leq
  \sigma \leq 4/5$. For fast rotating models, $\Omega / \sqrt{G
  \rho_c} > 0.25$, the $\rm r_s$-mode deviates from the simple
  slow-rotation relation~(\ref{eq:indr}), where $\gamma_{\varepsilon}
  = 1 / (1-\bar \varepsilon ) = 3$, because of the symmetry energy.
  In the right panel we show three $l=4, m=2$ superfluid inertial
  modes for $\sigma = -0.5$ and $\sigma = 0.5$. These results also
  show a clear dependence on the symmetry energy. It is worth noticing
  that, for the non-stratified models A the ordinary inertial modes
  are equal to the results for single fluid $N=1$~polytropes, where
  $N$ is the polytropic index.
\label{fig:rmodes}}
\end{center}
\end{figure}
%------------------------------------------------------------------------------%

%~~~~~~~~~~~~~~~~~~~~~~~~~~~~~~~~~~~~~~~~~~~~~~~~~
\section{Results} \label{sec:Res}
%~~~~~~~~~~~~~~~~~~~~~~~~~~~~~~~~~~~~~~~~~~~~~~~~~

In this section we present the result from the first ever
time-evolution study of perturbed, rapidly rotating, superfluid
stars. The main aim is to explore how fast rotation, symmetry energy
and entrainment affect the non-axisymmetric oscillation modes.

\subsection{The evolution code}

The evolution problem for
equations~(\ref{eq:dfdt})--(\ref{eq:dchipdt}) is intrinsically a three
dimensional problem. As already mentioned, the Fourier decomposition
of the azimuthal degree of freedom, identifying specific $m$-modes,
reduces the problem to two spatial dimensions.  In spherical
coordinates, we could then evolve this system of equations on a
two-dimensional grid based on the coordinates~$r$ and
$\theta$. However, instead of using $r$ we adopt a new radial
coordinate $x=x(r,\theta)$, fitted to surfaces of constant
pressure~\citep{2002MNRAS.334..933J,2008arXiv0807.3457P}. This leads
to an easier implementation of the surface boundary conditions.  As we
are working in the time domain, the various mode frequencies are
extracted by a Fast Fourier Transformation~(FFT) of the time evolved
perturbation variables.

The numerical code for superfluid neutron stars extends the single
fluid code developed by~\citet{2008arXiv0807.3457P}. In fact, with our
chosen variables the Euler equation~(\ref{eq:dfdt}) and the mass
conservation equation~(\ref{eq:drhodt}) are identical to the single
fluid case. We have thus extended the previous code by adding
equations~(\ref{eq:dDdt}) and~(\ref{eq:dchipdt}), together with the
appropriate boundary conditions.  The main elements of the code are
the use of a Mac-Cormack algorithm, a second order accurate numerical
scheme both in space and time, and the implementation of a
fourth-order Kreiss-Oliger numerical dissipation, which stabilizes the
simulations against spurious high frequency oscillations. The
performance of the final code is practically identical to that of the
single fluid code, see~\citet{2008arXiv0807.3457P} for details.

%------------------------------FIG. 2------------------------------------------%
\begin{figure}
\begin{center}
\includegraphics[height=75mm]{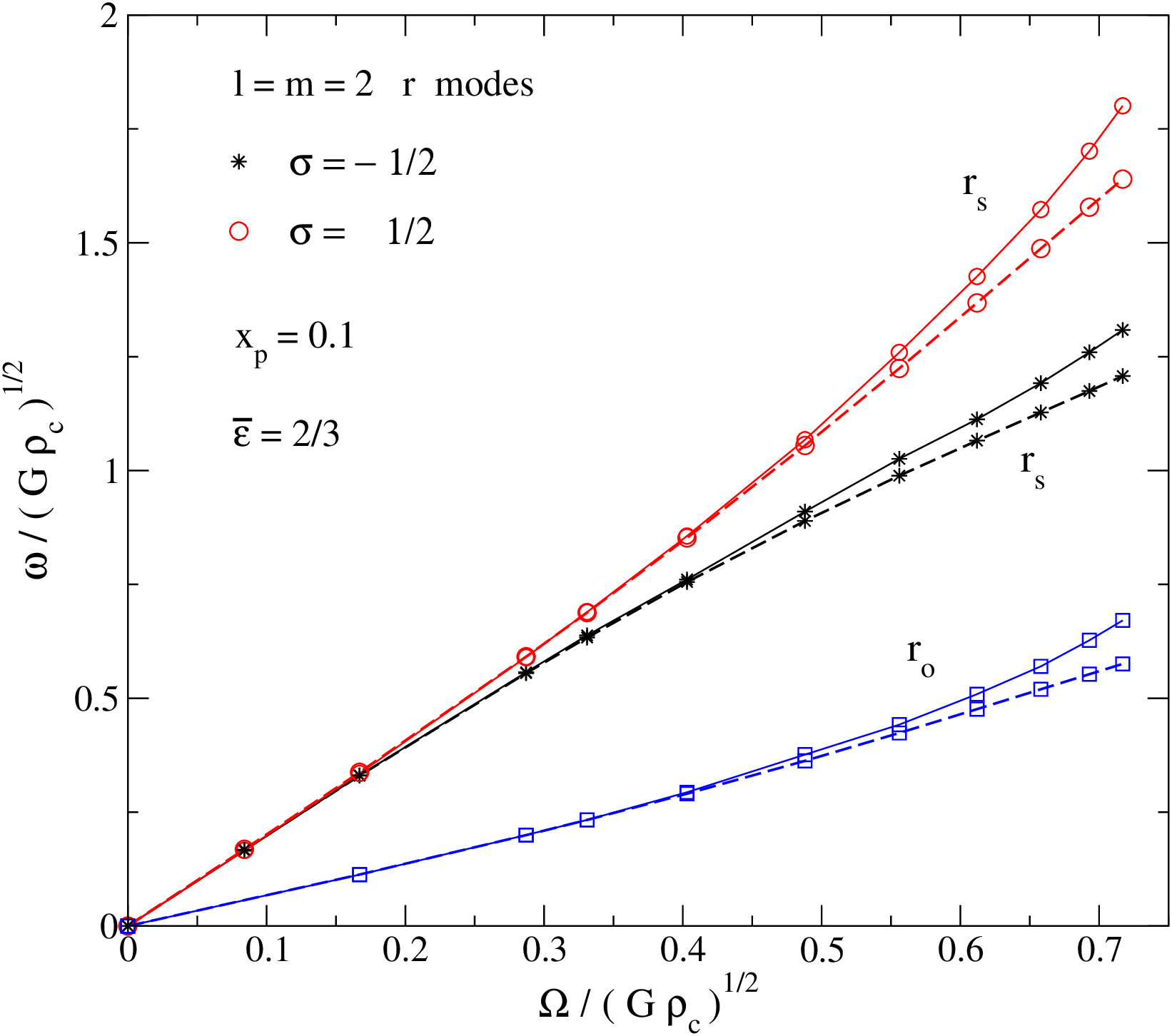}
\includegraphics[height=75mm]{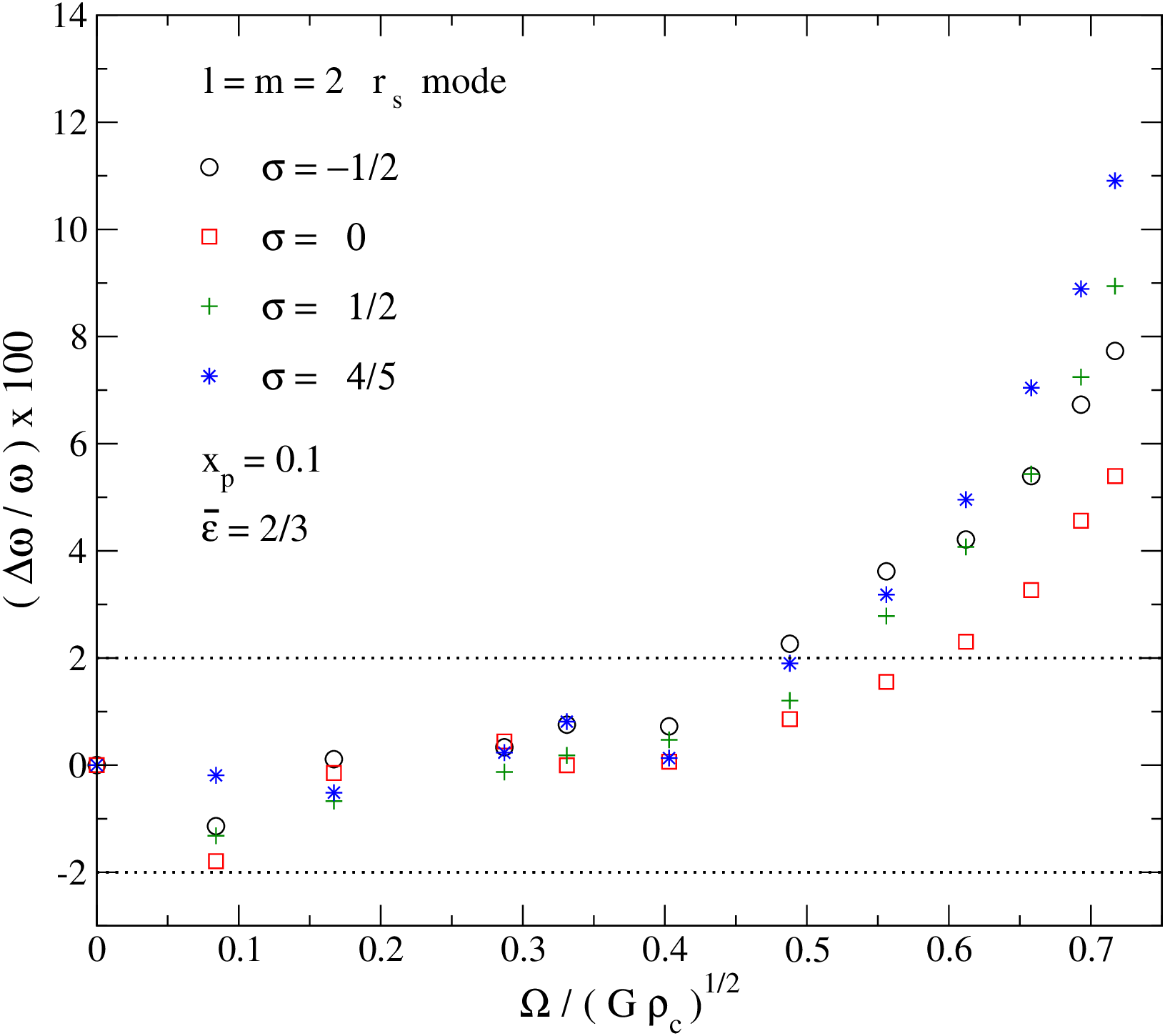}
\caption{In this figure we compare our rotating frame frequencies for
  the r-modes with the slow rotation results of ~\citet{2009arXiv0902.1149H}, 
  which include $\mathcal{O} \left( \Omega^3 \right)$ corrections.  In
  the left panel, we show the $l=m=2$ $\rm r_s$ and $\rm r_o$-mode
  frequencies for the rotating models A with $\sigma = \pm 1/2$,
  proton fraction $x_\p = 0.1$, and entrainment parameter~$ \bar
  \varepsilon = 2/3$. Our numerical results are shown as solid lines,
  while the frequencies of~\citet{2009arXiv0902.1149H} are represented by dashed
  lines. The right panel provides the relative error between the $\rm
  r_s$-mode frequencies calculated with the two methods, where the
  dotted lines denote the 2\% level. The results show that the calculated 
  frequencies agree to
  better than  2\% up to a stellar angular velocity of $\Omega
  / \sqrt{G \rho_c } \approx 0.4$.
\label{fig:rmodes-comp}}
\end{center}
\end{figure}
%------------------------------------------------------------------------------%

\subsection{Initial Data}

In a time domain study, the initial perturbations can be chosen to
excite specific parts of the spectrum.  Of course, a strict selection
of oscillation modes requires the determination of eigenfunctions to
be used as initial data. To achieve this, one would have to either
solve the frequency domain version of the perturbation
equations~(\ref{eq:dfdt})--(\ref{eq:dchipdt}) or perform an
eigenfunction recycling study of the time
evolutions~\citep{Stergioulas:2003ep, Dimmelmeier:2005zk}.  In this
work we use neither of these strategies. We are interested in a
multi-mode analysis where many oscillation modes are excited in each
simulation.  As discussed by~\cite{2008arXiv0807.3457P}, this is
easily achieved by the use of initial perturbations with an arbitrary
radial profile and an angular dependence appropriate for the general
class of eigenfunction. For the radial part we typically use a
Gaussian distribution. Meanwhile the angular functions are inspired by
slow-rotation results.

Type I parity perturbations are generally excited with the following
mass density and proton fraction perturbations:
\begin{eqnarray}
\delta \rho & = & A_0 \, \exp{\left[ - \left( \frac{r-r_0 }{ q  R\left( \theta \right)} \right) ^2 \right] }
           Y_{l l} \left(\theta,\phi\right) \, , \label{eq:fIP} \\
\delta \chi_\p & = & A_1  \, \exp{\left[ - \left( \frac{r-r_0 }{ q  R\left( \theta \right)} \right) ^2 \right] }
           Y_{l l} \left(\theta,\phi\right) \, , \label{eq:fIchiP}
\end{eqnarray}
where $A_0$ and $A_1$ are two arbitrary constants that determine the
initial values of~$\delta \rho$ and $\delta \chi_\p$ on the star's
surface. The stellar radius at polar angle $\theta$ is denoted by
$R\left( \theta \right)$, and the parameters $r_0$ and $q$
respectively determine the centre of the Gaussian profile and its
width. The $l=m$ spherical harmonic $Y_{l l} \left(\theta,\phi\right)$
approximates the typical angular behaviour of a polar mode in a
spherical star.  For simplicity, all other perturbation variables are
set to zero to complete the initial data.

For Type II parity perturbations we use the following initial data for
the vector fields~$\mtb{f}$ and~$\mtb{D}$:
\begin{eqnarray}
\mtb{f} & = & \rho \, \exp{\left[- \left( \frac{r-r_0 }{ q  R\left( \theta \right)} \right) ^2 \right] }
           Y_{l l}^B \left(\theta,\phi\right) \, , \\
\mtb{D} & = & x_\p \left( 1-x_\p \right) \rho \, \exp{\left[- \left( \frac{r-r_0 }{ q  R\left( \theta \right)} \right) ^2 \right] }
           Y_{l l}^B \left(\theta,\phi\right) \, ,
\end{eqnarray}
Here $Y_{l l}^B \left(\theta,\phi\right)$ is a magnetic spherical
harmonic~\citep{1980RvMP...52..299T}. The remaining perturbations are
set to vanish on the initial time slice.

%%%%%%%%%%%%%%%%%%%%%%%%%%%%%%%%%%%% SEC. %%%%%%%%%%%%%%%%%%%%%%%%%%%%%%%%%%%%%%%
\subsection{Inertial modes} \label{sec:Inerm}
%%%%%%%%%%%%%%%%%%%%%%%%%%%%%%%%%%%%%%%%%%%%%%%%%%%%%%%%%%%%%%%%%%%%%%%%%%%%%%%%%

Let us first consider the inertial modes which in a superfluid star
split (more or less clearly depending on the equation of state) into
ordinary and superfluid modes. In the first class, the perturbed fluid
elements of the two components oscillate in phase, whereas for the
superfluid modes they pulsate in counter-phase. As for single fluid
stars, each inertial mode can be classified by its parity as an
axial-led or polar-led inertial
mode~\citep{1999ApJ...521..764L}. Among the axial-led inertial modes,
the r-modes form a well defined sub-set. They are the only modes that
are purely axial in the slow-rotation limit. In stratified neutron
stars, only the co-moving r-mode exists. The counter-moving mode is no
longer purely axial, but acquires a polar component already at leading
order in the rotation~\citep{2009arXiv0902.1149H}.  In a non-stratified neutron
star, on the other hand, the ordinary and superfluid degrees of
freedom are completely decoupled and a purely axial counter-moving
r-mode exists~\citep{2008arXiv0812.3023A, 2009arXiv0902.1149H}.

Up to $\mathcal{O} \left( \Omega \right)$, the frequencies of the
superfluid and ordinary inertial modes are related according
to~\citep{2004MNRAS.348..625P}:
\begin{equation}
\omega_s \simeq \gamma_{\veps} \,  \omega_o \, ,  \label{eq:indr}
\end{equation}
where $\gamma_{\veps} \equiv (1-\bar{\veps})^{-1}$.  It makes sense,
as a first step, to investigate to what extent the inertial-mode
frequencies deviate from this simple scaling law in the case of rapid
rotation. To do this, we consider model A with fixed entrainment
parameter $\bar{\veps} = 2/3$ and proton fraction $x_\p = 0.1$. For
the symmetry energy term we consider four different values $\sigma = -
1/2, 0, 1/2 , 4/5$. The results are shown in
figure~\ref{fig:rmodes}. In the left panel we show the $l=m=2$
ordinary and superfluid r-modes. The ordinary $\rm r_o$ mode is
represented by a solid line and the expected frequency~(\ref{eq:indr})
of the superfluid $\rm r_{s}$ mode is also indicated.  Our results
show that the $\rm r_{s}$ mode frequencies for different values of
$\sigma$ agree well with equation~(\ref{eq:indr}) roughly up to a
stellar rotation $\Omega / \sqrt{ G \rho_c } \simeq 0.2$. For faster
rotation, the $\rm r_s$-mode frequency depends strongly on the
symmetry energy.

In order to confirm these results, we also considered the superfluid
r-modes within the slow-rotation approximation \citep{2009arXiv0902.1149H}. The
approximate results confirm that, even though one should expect the
counter-moving inertial modes to approach (\ref{eq:indr}) as $\sigma
\to 1$, the relation does not hold perfectly even in the extreme
limit.

Our results show that the $\rm r_s$-mode frequency remains closer to
the values expected from equation~(\ref{eq:indr}) for larger $\sigma$.
Similar behaviour is noted for other inertial modes. In the right
panel of figure~\ref{fig:rmodes}, we show three $l=4$, $m=2$ axial-led
inertial modes for $\sigma = \pm 1/2$. The dependence on the symmetry
energy term is distinguishable beyond $\Omega / \sqrt{ G \rho_c }
\approx 0.25$ also for these modes.  These results can be understood
from a local plane-wave analysis, see Appendix~\ref{sec:LA} for a
detailed discussion.

%------------------------------FIG. 3------------------------------------------%
\begin{figure}
\begin{center}
\includegraphics[height=75mm]{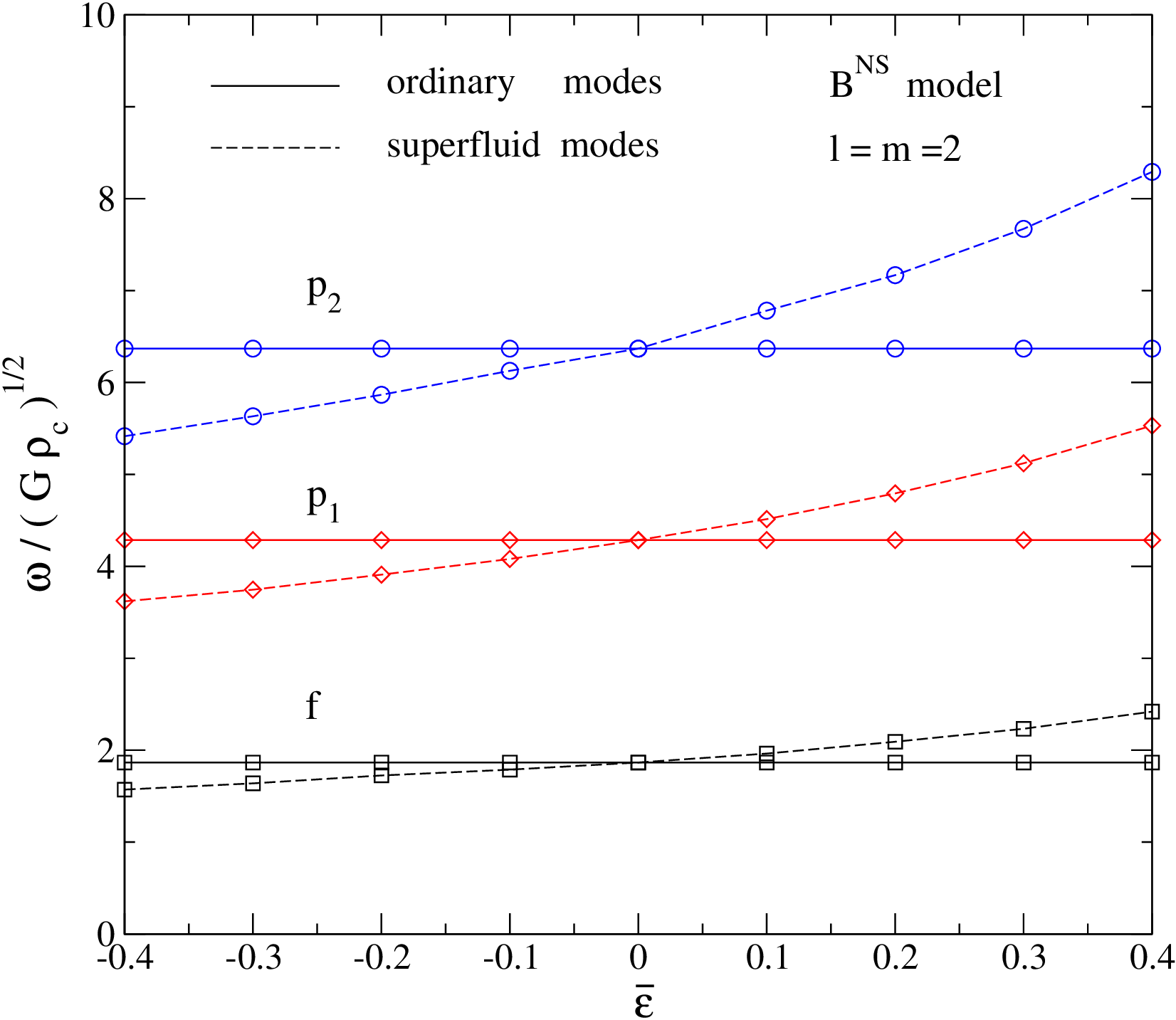}
\includegraphics[height=75mm]{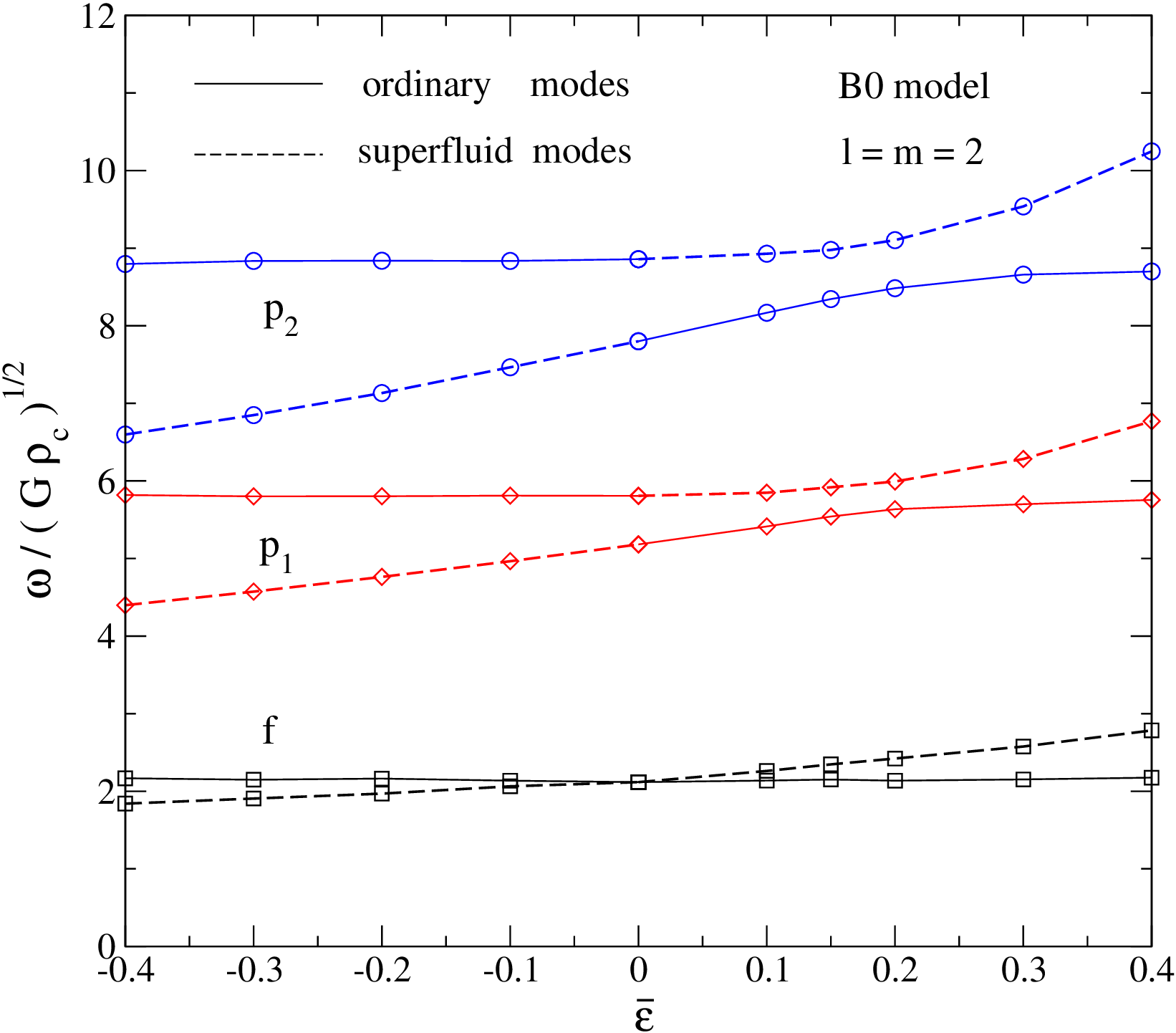}
\caption{This figure illustrates the dependence of the $l=m=2$
acoustic modes on the entrainment parameter $\bar \veps$ for two
non-rotating stellar models, the non-stratified $\rm B^{NS}$ model
(left panel) and the stratified B0 model (right panel). Solid and
dashed lines represent oscillation modes where neutrons and protons
move, respectively, in phase (ordinary modes) and counter-phase
(superfluid modes). The spectrum of the B0 model shows some avoided
crossings. Avoided crossings are expected in a stratified neutron star
as the co- and counter-moving degrees of freedom are coupled.
\label{fig:1PR-eps}}
\end{center}
\end{figure}
%------------------------------------------------------------------------------%

We can also compare our results to the superfluid r-mode frequencies
calculated by~\citet{2009arXiv0902.1149H}, who worked in the slow-rotation
approximation keeping terms up to $\mathcal{O} \left( \Omega ^ 3
\right)$.  The $l=m=2$ $\rm r_s$ frequency is then given by
\begin{equation}
\omega_\s = c_0 \, \Omega + c_2 \, \Omega^3 \, , \label{eq:om-slrot}
\end{equation}
where $c_0$ and $c_2$ are constants that depend on the stellar model
and the multipole of the modes. The first coefficient has the well
known analytical expression:
\begin{equation}
c_0 = \gamma_{\veps} \, \frac{ 2 m } { l \left( l + 1 \right) } \, ,
\end{equation}
while $c_2$ is given in closed form by
\citet{2009arXiv0902.1149H}. For the $l=m=2$ $\rm r_s$-mode of a model
A star with $\bar{\veps} = 2/3$ and $x_\p = 0.1$, we have $c_0 = 2$
and the values for $c_2$ given in Table~\ref{tab:Brym-coeff}. In
figure~\ref{fig:rmodes-comp}, we compare the $\rm r_s$-mode
frequencies extracted from our evolutions to those determined
analytically by~\citet{2009arXiv0902.1149H} for different values of
the symmetry energy.  In the left panel, we show how the $\rm
r_s$-mode frequency depends on the star's rotation for two selected
models with $\sigma = \pm 1/2$. The right panel shows the relative
error between the frequencies obtained with the two methods. The
agreement is better than three percent up to $\Omega / \sqrt{ G \rho_c
} \approx 0.55$, which corresponds to a rapidly rotating star with
$\Omega / \Omega_K \approx 0.77$~(see
Table~\ref{tab:back-models}). For faster rotation, the errors become
larger and reach eleven per cent near the mass shedding limit.  In
order to improve the accuracy, the slow-rotation analysis would need
to consider higher order corrections. This may be prohibitively
difficult.

%------------------------------TAB. 3------------------------------------------%
\begin{table}
\begin{center}
\caption{\label{tab:Brym-coeff} The coefficient $c_2$ required in
equation~(\ref{eq:om-slrot}) for the $l=m=2$ superfluid r-mode. The
values of $c_2$ are given in the units used
in~\citet{2009arXiv0902.1149H}, i.e.  $ G M / R^3$.  Here we have used
the mass $M$ and the radius $R$ for the non-rotating model A0, with
$\bar{\veps} = 2/3$ and $x_\p = 0.1$. }
\begin{tabular}{ c c }
\hline $\sigma $ &  $ c_2 $  \\
\hline
\hspace{-0.3cm} $-0.5$ &  \hspace{-0.3cm} $-1.052$    \\
0.0   & 0.452      \\
0.5   & 0.956      \\
0.8   & 1.123      \\
\hline
\end{tabular}
\end{center}
\end{table}
%------------------------------------------------------------------------------%

The dependence of the superfluid inertial modes on the proton fraction
is very weak for model A neutron stars. That this should be expected
can be seen from equation~(\ref{eq:LAp}). To confirm this we have
evolved the fast rotating model A8 with $x_\p = 0.01$. The frequencies
for this model differ by less than one percent from the $x_\p = 0.1$
case.

%%%%%%%%%%%%%%%%%%%%%%%%%%%%%%%%%%%%%% SEC %%%%%%%%%%%%%%%%%%%%%%%%%%%%%%%%
\subsection{Acoustic modes}
%%%%%%%%%%%%%%%%%%%%%%%%%%%%%%%%%%%%%%%%%%%%%%%%%%%%%%%%%%%%%%%%%%%%%%%%%%%

Let us now study the effects of entrainment and symmetry energy on the
fundamental and pressure modes of a superfluid star.  The acoustic
mode spectrum is characterized by the usual two classes of ordinary
(co-moving) and superfluid (counter-moving) modes.  We will focus on
the quadrupole $l=m=2$ modes. These are the modes that tend to be the
most relevant for gravitational-wave studies.

First we consider the oscillations of two non-rotating equilibrium
configurations, the non-stratified $\rm B^{NS}$ model and the
stratified B0 model described in Section~\ref{sec:BGEoS_2}.  In
figure~\ref{fig:1PR-eps}, we show the variation of the fundamental
quadrupole mode and the first two quadrupole pressure modes
with the entrainment parameter $\bar{\veps}$.  In order to determine
the co- and counter-moving character of a mode, we have first
reconstructed the time variation of the component velocities
$\mtb{v_\x}$ from the primary dynamical variables~$\mtb{f}$
and~$\mtb{D}$. Post processing the numerical evolution data, we have
determined the velocity components using the eigenfunction extraction
code developed by~\citet{Stergioulas:2003ep}
and~\cite{Dimmelmeier:2005zk}.  For any mode, we have then compared
the velocity eigenfunctions of the two fluid components and determined
whether they oscillate in phase or counter-phase. In
figure~\ref{fig:1PR-eps}, the solid lines represent modes that
oscillate in phase, while dashed lines correspond to modes that
pulsate in counter-phase.  In absence of composition gradients (model
$\rm B^{NS}$ in the left panel of figure~\ref{fig:1PR-eps}), the
co-moving and counter-moving degrees of freedom are completely
decoupled and the spectrum does not exhibit any interaction between
ordinary and superfluid modes. They are actually related by the simple
expression, cf.~(\ref{eq:LAp}),
\begin{equation}
\omega_\s \simeq \sqrt{\gamma_{\veps}} \, \omega_\ord \,
. \label{eq:fsNR}
\end{equation}

The spectrum of stratified superfluid stars is more interesting. From
the right panel of figure~\ref{fig:1PR-eps}, we note that the coupling
between ``ordinary'' and ``superfluid'' perturbations generates
avoided crossings, where the oscillation phase of the mode changes at
(roughly) $\bar{\veps} = 0$.  This behaviour is more evident in the
ordinary and superfluid pressure modes. From our data there does not
appear to be an avoided crossing for the fundamental mode. However,
the region where this crossing should take place is difficult to
resolve with our evolutions. We have tried different initial data sets
but only managed to find a single f-mode at $\bar \veps = 0$.
Nonradial oscillations of the $\rm B^{NS}$ and B0 models have already
been studied by~\citet{2002A&A...393..949P}, although not within the
Cowling approximation. A direct comparison with their results is not
possible since the Cowling approximation can introduce a 15-20$\%$
error in the f-mode frequencies. However, the qualitative behaviour of
the acoustic spectrum in the two studies is clearly similar.

%------------------------------FIG. 4------------------------------------------%
\begin{figure}
\begin{center}
\includegraphics[height=75mm]{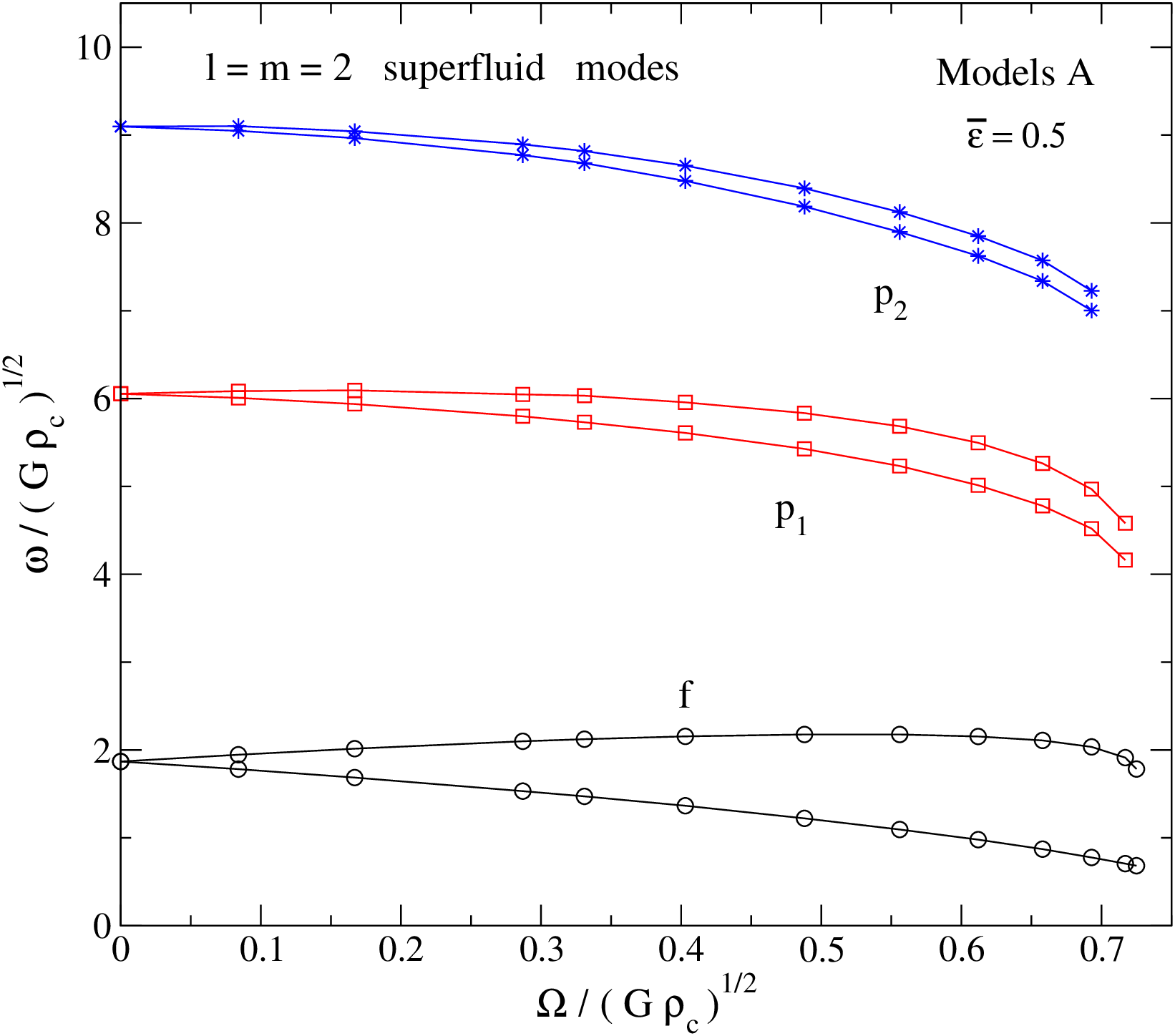}
\includegraphics[height=75mm]{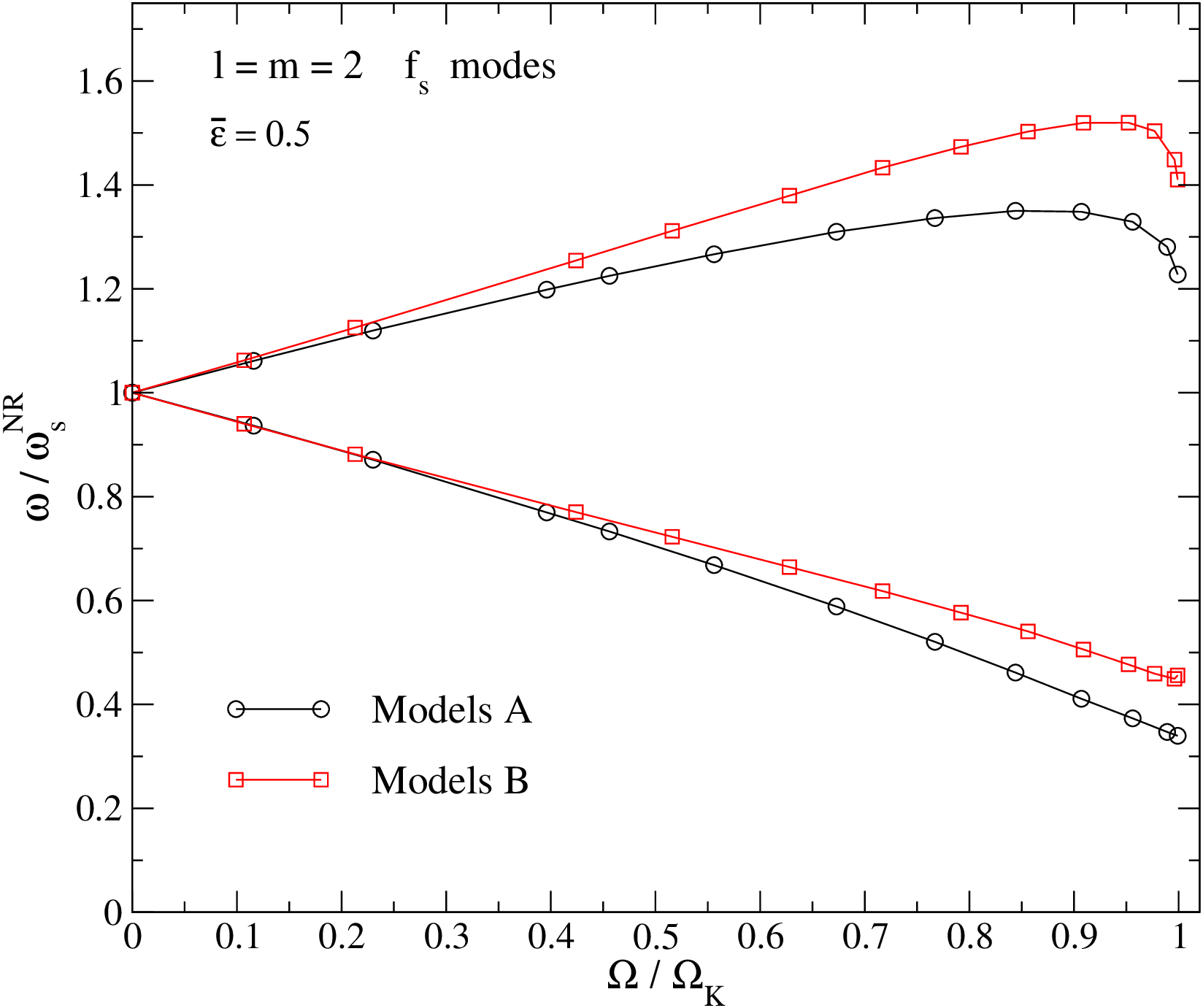}
\caption{This figure shows the rotational splitting of the $l=m=2$
non-axisymmetric modes (in the reference frame of the rotating star)
for the two sequences of rotating models A and B. In the left panel,
we show the dimensionless frequencies of the superfluid f-mode and the
first two counter-moving p-modes for models A with $\sigma=0$ and
$\bar \veps = 0.5$. In the right panel, we compare the superfluid
f-modes of these A models with the models B having $\bar \veps =
0.5$. In this panel, the star's angular velocity is normalized (the
horizontal axis) with the mass-shedding limit $\Omega/\Omega_K$, while
the mode-frequency is divided by the superfluid f-mode frequency of
the non-rotating star $\omega_\s^{\rm{NS}}$ (the vertical axis).
\label{fig:f2mode-modAB}}
\end{center}
\end{figure}
%------------------------------------------------------------------------------%

In order to investigate the effect of entrainment on the rotational
splitting of the non-axisymmetric acoustic modes we consider a
sequence of A models with $\sigma=0$ and models B. In
figure~\ref{fig:f2mode-modAB}, we show results corresponding to the
$l=m=2$ acoustic modes in the $\bar \veps = 0.5$ case. For the generic
initial data given by equations~(\ref{eq:fIP})--(\ref{eq:fIchiP}) many
oscillation modes are excited in the numerical simulations. By
studying their eigenfunctions we can track individual modes from the
non-rotating model up to the mass shedding limit. An example is
illustrated in the left panel of figure~\ref{fig:f2mode-modAB}, where
the rotating frame frequencies of the superfluid f-mode and the first
two p-modes are shown for the A models.  The counter-moving f$_\s$
mode exhibits a larger rotational splitting compared to the two
pressure modes.  In the right panel of figure~\ref{fig:f2mode-modAB},
we compare instead the normalized frequencies of the superfluid
f-mode, namely $\omega / \omega^{\rm{NR}}_{\s}$, for models A and
B. The horizontal-axis of this figure represents the stellar rotation
divided by the maximal angular velocity $\Omega_K$. The results in the
figure show that, even though the two classes of models have the same
value of the entrainment parameter, the non-axisymmetric splitting of
the f$_\s$ mode is quite different in the two cases. This is
particularly clear in the rapid rotation regime. Meanwhile, the two
$\rm f_\s$ modes have very similar frequencies for $\Omega \lesssim
0.25~\Omega_K$.

\begin{figure}
\begin{center}
\includegraphics[height=75mm]{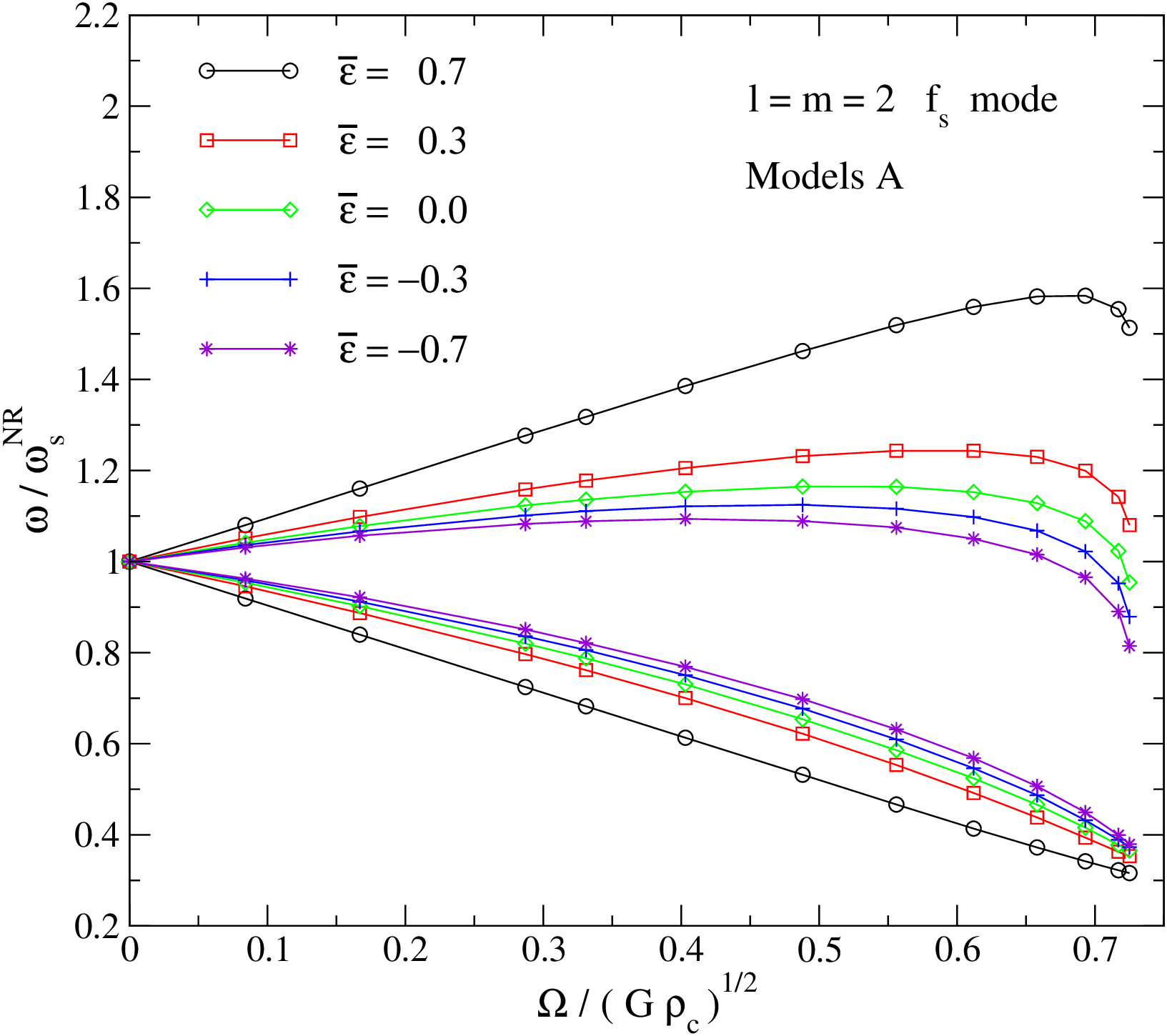}
\includegraphics[height=75mm]{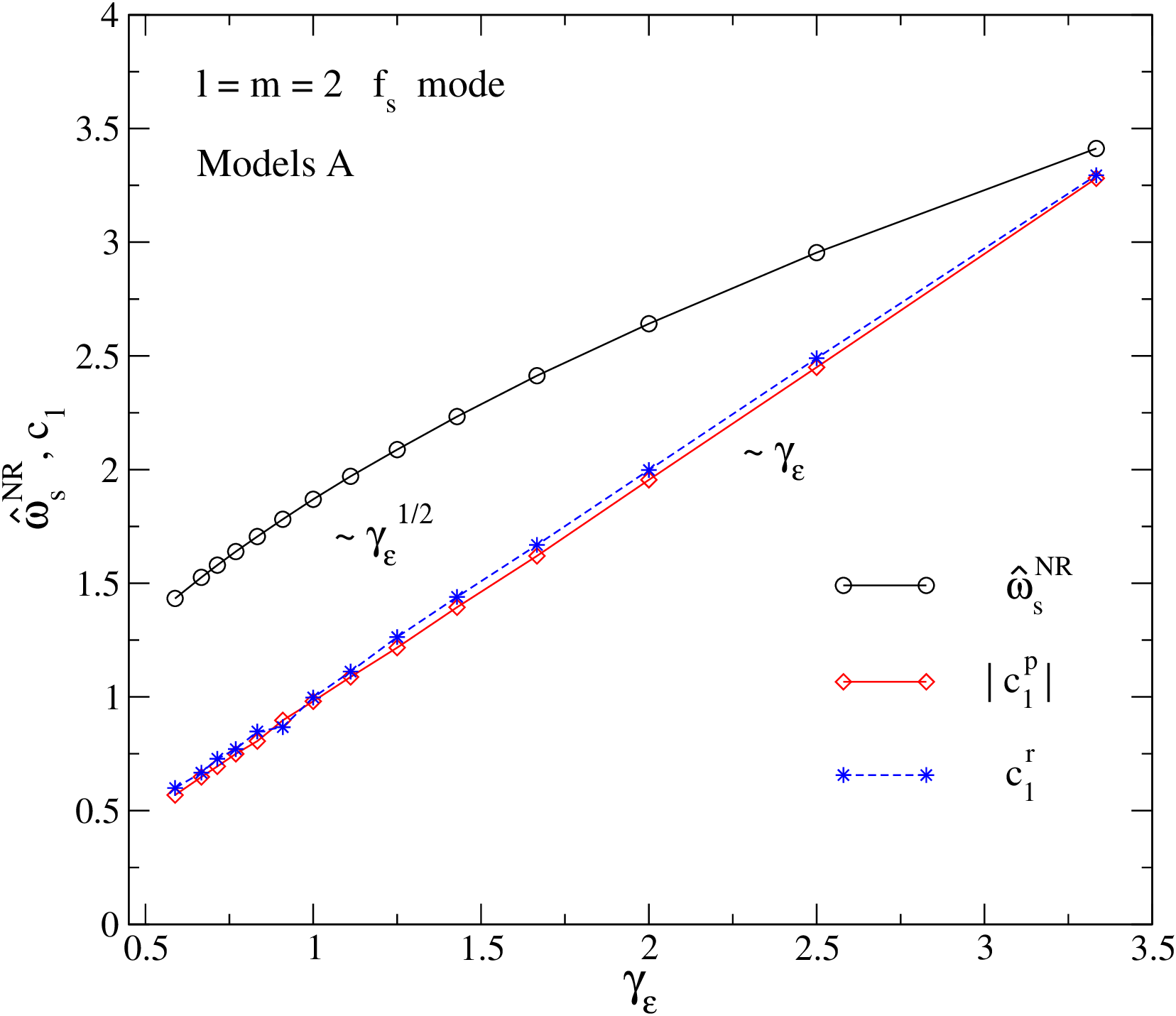}
\caption{The effect of the entrainment parameter $\bar \veps$ on the
rotational splitting is shown in this figure. For the sequence of
stellar models A with $\sigma = 0$, the left panel displays the
rotating frame frequencies of the $l=m=2$ superfluid modes for several
values of $\bar \veps$. The mode-frequencies are normalized with the
superfluid f-mode frequencies of the non-rotating star
$\omega_{\s}^{\rm{NR}}$. In the right panel, we show how the
dimensionless frequency $\hat{\omega}_{\s}^{\rm{NR}} = \omega_\s /
\sqrt{ G \rho_c }$ and the two splitting parameters $c_1^{r}$ and
$c_1^{p}$ of equation~(\ref{eq:fit-f2}) depend on the entrainment
parameter $\gamma_{\veps} = \left( 1 - \bar \veps \right)^{-1}$. The
superscripts $r$ and $p$ for the parameters $c_1^{r}$ and $c_1^{p}$
denote retrograde and prograde modes, respectively. The quantities
$c_1^{r}$ and $c_1^{p}$ have been determined by a fit to the
superfluid f-mode frequencies for models A up to $\Omega \lesssim
0.375~\Omega_K$.
\label{fig:Amodel-f2-split}}
\end{center}
\end{figure}

It is instructive to explore the slow-rotation regime to see if there
is a simple relation between the $\rm f_\s$ mode frequencies and the
entrainment. To this end, we determine the $\rm f_\s$ mode
frequencies for models A and B with $-0.7 \leq \bar \veps \leq
0.7$. In the left panel of figure~\ref{fig:Amodel-f2-split}, the
variation of the $\rm f_\s$ mode frequency with $\bar \veps$ is shown
for models A. It is useful to recall that these models have $\sigma=0$
by construction.  All modes have been normalized to the f-mode
frequency of the non-rotating star. The rotational splitting of the
modes strongly depends on the parameter $\bar \veps$.  In order to
quantify the effect, we approximate the dimensionless mode-frequency
$\hat{\omega}_\s = \omega_\s / \sqrt{ G \rho_c }$ (see
Appendix~\ref{sec:LA}) by a second order polynomial in $\hat{\Omega}
= \Omega / \sqrt{ G \rho_c }$,
\begin{equation}
\hat{\omega}_\s = \hat{\omega}_\s^{\rm{NR}} + c_1 \left( \veps, \sigma, m \right)
\hat{\Omega} + c_2 \left( \veps, \sigma, m \right) \hat{\Omega}^2 + \mathcal{O}
\left( \hat{\Omega}^3 \right) \, , \label{eq:fit-f2}
\end{equation}
where $\hat{\omega}^{\rm{NR}}_\s = \omega^{\rm{NR}}_\s/
\sqrt{G\rho_c}$ is the superfluid mode frequency of the non-rotating
star, while $c_1$ and~$c_2$ are two parameters that we fit to the
numerical data.  In general, these parameters are functions of the
entrainment, the symmetry energy and the multipole $m$ of the mode.
Since this is inherently a slow-rotation approximation, we will only
consider rotating models with $\Omega \lesssim 0.375~\Omega_K$.  An
accurate description of the spectrum for more rapidly rotating models
would require higher order fitting functions. In the right panel of
figure~\ref{fig:Amodel-f2-split}, we show the dependence of
$\hat{\omega}^{\rm{NR}}_\s$ and $c_1$ on
\begin{equation}
\gamma_{\varepsilon}= \left( 1 - \bar \varepsilon \right)
^{-1} \, ,
\end{equation}
for Models A. We have defined $c_1^{r}$ and $c_1^{p}$ for retrograde
and prograde modes, respectively. If we assume that a perturbation
variable is proportional to $e^{i\left( \omega t + m \phi \right)}$,
modes with positive (negative) $m$ move retrograde (prograde) with
respect to the stellar rotation.

For non-stratified and non-rotating stellar models,
equation~(\ref{eq:LAp}) suggests that the superfluid and ordinary
f-mode frequencies are related according to
\begin{equation}
\hat{\omega}_\s^{\rm{NR}} = \sqrt{\gamma_{\veps}} \, \chi (\sigma,
x_\p) \, \hat{\omega} _{\ord}^{\rm{NS}} \, , \label{eq:NR-omS}
\end{equation}
where $\chi$ is a function of $\sigma$ and $x_\p$ that, in general,
depends on the EoS. For models A, this function is given explicitly by
\begin{equation}
\chi ^2 \equiv \frac{\left( 1 + \sigma \right) \left( 1-x_\p \right)}{
  1 - \left(1+\sigma \right) x_\p } \, .
\end{equation}
Since we consider a sequence of models with $\sigma = 0$ it follows
that $\chi=1$. We test the accuracy of equation~(\ref{eq:NR-omS})
against the f-mode frequencies for models A shown in
figure~\ref{fig:Amodel-f2-split}. Fitting our numerical data to a
function of form
\begin{equation}
a \gamma_{\veps} ^{b} \, , \label{eq:fit-Psi1}
\end{equation}
we obtain the values for $a$ and $b$ listed in
Table~\ref{tab:fit-par}. These results are in very good agreement with
the analytical formula~(\ref{eq:NR-omS}). In fact, the ordinary f-mode
frequency determined from our code is $\hat{\omega}_{\ord} = 1.867$,
which agrees to better than $0.06\%$ with the fitted value.
Equation~(\ref{eq:NR-omS}) also provides an accurate representation
for the f-mode of the B models. In this case, the numerical evolutions
lead to $\hat{\omega}_{\ord} = 2.1598$, which is within the error bar
of the fitted result, see Table~\ref{tab:fit-par}.  It should be noted
that we do not have a simple analytic expression for $\chi$ in the
case of the B models. Our results suggest that $\chi$ is only weakly
dependent on the entrainment also for this equation of state.

%------------------------------FIG. 5------------------------------------------%
%------------------------------------------------------------------------------%

Let us now consider the rotational corrections to the counter-moving
f-mode frequency, cf.  equation~(\ref{eq:fit-f2}). We find that we can
still use the fitting function~(\ref{eq:fit-Psi1}) for the parameter
$c_1$.  This leads to the results given in
Table~\ref{tab:fit-par}. These results suggest that the rotational
splitting parameter $c_1$ depends linearly on the entrainment
parameter $\gamma_{\veps}$ for both models A and B. This is also clear
from the results in Figure~\ref{fig:Amodel-f2-split}.  To describe the
quantities $c_2^r$ and $c_2^p$, we instead used the following fitting
function:
\begin{equation}
\frac{ a + b \, \gamma_{\veps}}{\hat{\omega}^{\rm{NR}}_\s} \, .
\label{eq:fit-Psi2}
\end{equation}
The results for $a$ and $b$ are given in Table~\ref{tab:fit-par}.
 Despite the fact that models B are stratified, the $\rm f_\s$ mode
 seems to maintain the same dependence on the entrainment as for the
 non-stratified models A in the slow rotation regime.
 
\begin{figure}
\begin{center}
\includegraphics[height=75mm]{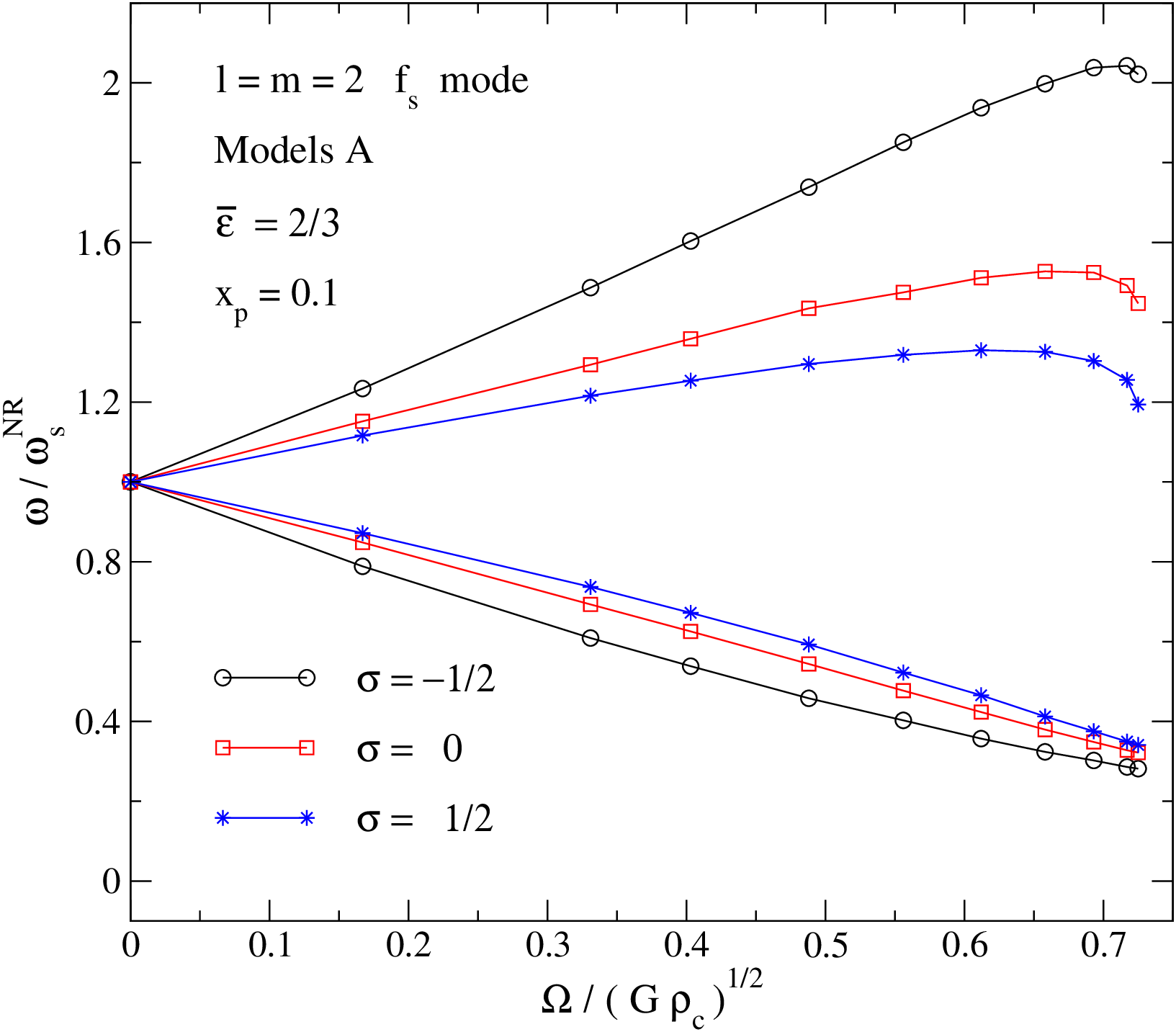}
\caption{This figure shows the effect of the symmetry energy on the
 rotational splitting of the quadrupole superfluid f-mode. The data in
 the figure corresponds to stellar models A with constant proton
 fraction $x_\p =0.1$ and entrainment $\bar \veps = 2/3$. The symmetry
 energy takes the three values $\sigma = -0.5,0,0.5$.
\label{fig:Amodel-f2-sigma}}
\end{center}
\end{figure}

 Building on this, let us try to understand the dependence on the
symmetry energy.  To do this we consider the sequence of rotating
stellar models A and take the symmetry energy in the range $ -1/2 \leq
\sigma \leq 1/2$. Meanwhile the proton fraction and the entrainment
are kept constant at $x_\p =0.1$ and~$\bar \veps = 2/3$,
respectively. The $\rm f_\s$ mode dependence on the stellar spin (for
some of these models) is shown in figure~\ref{fig:Amodel-f2-sigma}.
In the figure, we have normalized the mode-frequencies to the f-mode
frequency of the non-rotating model. The results show that the
rotational splitting of the $l=m=2$ superfluid f-mode decreases for
larger values of the symmetry energy. In the slow rotation regime, we
can study this effect by fitting the numerical data using
equation~(\ref{eq:fit-f2}). Adding the dependence on the symmetry
energy to our previous results, we now use the fitting function
\begin{equation}
\hat{\omega}_\s = \sqrt{\gamma_{\veps}} \, \chi(\sigma, x_\p) \,
\hat{\omega} _{\ord}^{\rm{NS}} + \gamma_{\veps} \, \tilde{c}_1 \left(
\sigma, m \right) \hat{\Omega} + \tilde{c}_2 \left( \veps, \sigma, m
\right) \hat{\Omega}^2 + \mathcal{O} \left( \hat{\Omega}^3 \right) \,
,
\label{eq:fit-f2-sigma}
\end{equation}
where $\tilde{c}_2$ is in general different from $c_2$, and we have
assumed that $\tilde c_1 $ is a function of only $\sigma$ and $m$.
First we consider the non-rotating case, using the ordinary frequency
of the f-mode $\hat{\omega} _{\ord}^{\rm{NS}}$ as a fitting
parameter. The fitted result agrees with the numerical value
determined from the simulations to better than $0.04\%$. Empirically
we find that the parameters $\tilde{c}_1$ can be approximated by the
function:
\begin{equation}
 \frac{ \gamma_{\veps} \, \tilde{a} \,
\chi ^{\tilde{b}} }{\hat{\omega}_\s^{\rm{NR}} } \, ,
\label{eq:fit-chi}
\end{equation}
The determined values for $\tilde a$ and $ \tilde b$ for model A are
given in Table~\ref{tab:fit-par-sig}.  We find that the parameter
$\tilde{c}_2$ assumes less regular values. They can be fitted with a
quadratic polynomial in $\sigma$, but the fits are not very accurate.
Hence, we do not provide any of those results here.  However, as our
main aim was to provide an approximate description of the rotational
splitting of the quadrupole superfluid f-mode, reliable values for the
coefficient $\tilde{c}_1$ should be sufficient.

%------------------------------FIG. 6------------------------------------------%
%------------------------------------------------------------------------------%

%------------------------------TAB. 4------------------------------------------%
\begin{table}
\begin{center}
\caption{\label{tab:fit-par} This table provides the values of the
  fitting parameters~$a\pm\Delta a$ and~$b\pm\Delta b$ from
  equations~(\ref{eq:fit-Psi1}) and~(\ref{eq:fit-Psi2}) for the
  superfluid f-mode of the non-rotating star,
  i.e. $\hat{\omega}^{\rm{NR}}_\s$, and the splitting coefficients
  $c_k^{r}$ and $c_k^{p}$ with $k=1,2$ for the retro- and pro-grade
  $\rm f_\s$ modes. In the fits we have used the stellar models A and
  B with~$\Omega \lesssim 0.375~\Omega_K$.}
\begin{tabular}{ l  c c c c c }
\hline                                     &  Models    &  $a$   &  $\Delta a$       &  $b$    & $\Delta b$  \\
\hline
\vspace{0.1cm} $\hat{\omega}_\s^{\rm{NR}}$ & A          & 1.8686 &  $2.1\times10^{-4}$  &  0.5000   & $1.5\times10^{-4}$ \\
\vspace{0.1cm} $c_1^r$                     & A          & 1.0034 &  $4.7\times10^{-3}$  &  0.9896   & $5.4\times10^{-3}$ \\
\vspace{0.1cm} $c_1^p$                     & A          & \hspace{-0.25cm} -0.9744 &  $1.7\times10^{-3}$  &  1.0072   & $2.1\times10^{-3}$ \\
\vspace{0.1cm} $c_2^r$                     & A          &  \hspace{-0.25cm} -0.5306 &  $1.7\times10^{-2}$  &  0.1513 & $1.1\times10^{-2}$ \\
$c_2^p$                                    & A          & \hspace{-0.25cm}  -0.5203 &  $4.4\times10^{-3}$  &  0.1570 & $2.8\times10^{-3}$ \\
\hline
\vspace{0.1cm} $\hat{\omega}_\s^{\rm{NR}}$ & B          & 2.1597 &  $8.6\times10^{-4}$  &  0.4987   & $1.0\times10^{-3}$ \\
\vspace{0.1cm} $c_1^r$                     & B          & 1.0091 &  $3.4\times10^{-3}$  &  0.9998   & $7.6\times10^{-3}$ \\
\vspace{0.1cm} $c_1^p$                     & B          &\hspace{-0.16cm}-0.9837 &  $3.8\times10^{-3}$  &  1.0059   & $8.7\times10^{-3}$ \\
\vspace{0.1cm} $c_2^r$                     & B          & \hspace{-0.16cm}-0.1421 &  $7.8\times10^{-3}$ &\hspace{-0.16cm} -0.0998  & $6.6\times10^{-3}$ \\
               $c_2^p$                     & B          &\hspace{-0.16cm}-0.1461 &  $1.2\times10^{-2}$  &  0.1106   & $1.1\times10^{-2}$ \\
\hline
\end{tabular}
\end{center}
\end{table}
%------------------------------------------------------------------------------%

%------------------------------TAB. 5------------------------------------------%
\begin{table}
\begin{center}
\caption{\label{tab:fit-par-sig} This table provides the values of the
  fitting parameters~$\tilde{a}\pm\Delta \tilde{a}$
  and~$\tilde{b}\pm\Delta \tilde{b}$ from
  equation~(\ref{eq:fit-chi}). They describe the splitting functions
  $\tilde{c}_1^{r}$ and $\tilde{c}_1^{p}$ for the $l=m=2$ superfluid
  f-mode.  The labels $r$ and $p$ denote retro- and pro-grade $\rm
  f_\s$ modes, respectively. In the fits we have used results for
  models A with~$\Omega \lesssim 0.375~\Omega_K$.}
\begin{tabular}{ l  c c c c c }
\hline                                     &  Models    &  $\tilde a$   &  $\Delta \tilde a$       &  $\tilde b$    & $\Delta \tilde b$  \\
\hline
\vspace{0.1cm} $\tilde{c}_1^r$             & A          & 0.3721 &  $1.1\times10^{-3}$  &  -0.985   & $1.5\times10^{-2}$ \\
$\tilde{c}_1^p$                            & A          & \hspace{-0.25cm} -0.3045 &  $1.5\times10^{-3}$  &  -1.024   & $2.0\times10^{-2}$ \\
\hline
\end{tabular}
\end{center}
\end{table}
%------------------------------------------------------------------------------%

%%%%%%%%%%%%%%%%%%%%%%%%  SEC: Conclusions  %%%%%%%%%%%%%%%%%%%%%%%%%%%%%%%
\section{Conclusions\label{conclusions}} \label{sec:concl}
%%%%%%%%%%%%%%%%%%%%%%%%%%%%%%%%%%%%%%%%%%%%%%%%%%%%%%%%%

In this paper we have considered, for the first time, the oscillations
of a superfluid neutron star as an initial-value problem. Using time
evolutions of the relevant linearised equations we studied
non-axisymmetric oscillations of rapidly rotating superfluid neutron
stars. We considered perturbations of axisymmetric background
configurations in Newtonian gravity and accounted for the presence of
superfluid components via the standard two-fluid model.  Within the
Cowling approximation, we were able to carry out evolutions for
uniformly rotating stars up to the mass-shedding limit.  Our results
represent the first detailed analysis of superfluid neutron star
oscillations in the fast rotation regime, where the star is
significantly deformed by the centrifugal force.

For simplicity, we focused on background models such that the two
fluids (superfluid neutrons and protons) co-rotate, are in
$\beta$-equilibrium and coexist in all the volume of the star. Two
different analytical model equations of state were considered. The
models were chosen to represent relatively simple generalizations of
single fluid, polytropic stars. We investigated the effects of
entrainment, rotation and symmetry energy on various non-radial
oscillation modes of these models. Our results show that entrainment
and symmetry energy can have a significant effect on the rotational
splitting of non-axisymmetric modes. In particular, the symmetry
energy modifies the inertial mode frequencies considerably in the
regime of fast rotation.

The perturbative time-evolution framework provides a useful tool that
should allow us to consider more realistic (and by necessity
complicated) neutron star models in the future.  The clear advantage
over frequency-domain studies is that it is straightforward to study
oscillations corresponding to eigenfunctions with a complex set of
rotational couplings. This is particularly useful in the rapid
rotation regime.  The obvious downside is that time evolutions can
never provide the ``complete" mode spectrum of the star.  Initial data
has to be chosen in such a way that the oscillations of interest are
excited at a significant level. It is difficult to, without prior
knowledge, find initial data that excites only a few modes. In many
cases this is, however, less relevant. The main question is if one can
extract accurate information regarding the nature of the star's
oscillations from the numerical data. The results we have presented
demonstrate that this is, undoubtedly, the case. Hence, it is relevant
to develop the perturbative evolution framework further. We are
currently considering more general background models, with a relative
velocity between the two fluid components. We are also adding the
dissipative coupling associated with mutual friction to the code.
Once these features are incorporated we will be able to consider
(obviously still at a basic level) the dynamics associated with the
superfluid two-stream instability \citep{2stream1,2stream2} and the
possible relation with pulsar glitches~\citep[see][for a recent
discussion]{2008arXiv0806.3664G}.  This is a very exciting prospect.
Looking further ahead, we would like to add layering to the stellar
model by introducing both an elastic crust and distinct
superfluid/normal regions. There are challenges associated with these
aspects, but there is no reason why these developments should be
prohibitively difficult.

%%%%%%%%%%%%%%%%%%%%%%%%  ACKNOWLEDGEMENTS  %%%%%%%%%%%%%%%%%%%%%%%%%%%%%%%%%%%%
\section*{Acknowledgements}
This work was supported by STFC through grant number PP/E001025/1.

%%%%%%%%%%%%%%%%%%%%%%%%%%%%%%%  APPENDICES  %%%%%%%%%%%%%%%%%%%%%%%%%%%%%%%%%%%
\appendix

%~~~~~~~~~~~~~~~~~~~~~~~~~~~~~~~~~~~~~~~~~~~~~~~~~
\section{Local Analysis} \label{sec:LA}
%~~~~~~~~~~~~~~~~~~~~~~~~~~~~~~~~~~~~~~~~~~~~~~~~~

In this Appendix we carry out a local analysis of the perturbation
equations~(\ref{eq:dfdt})--(\ref{eq:dchipdt}). The aim is to obtain
understand the superfluid oscillation spectrum and its dependence on
parameters like the proton fraction, the entrainment and the symmetry
energy. Since we are only interested in a qualitative picture we focus
on the non-stratified $N=1$~polytropic model A~(see
Section~\ref{sec:BGEoS_1}). For this sequence of models, the ordinary
and superfluid perturbation variables are decoupled. We consider only
the ``superfluid'' perturbations as the dispersion relations for the
ordinary modes are well known, see for
example~\citep{1989nos..book.....U}.
For model A the superfluid perturbation equations~(\ref{eq:dDdt}) and
(\ref{eq:dchipdt}) can be written
\begin{eqnarray}
\gamma_{\varepsilon}^{-1} \,  \partial_t \mtb{D}    & = &  -  \alpha^2  \nabla \delta \chi _\p
                            -   2 \mtb{\Omega} \times \mtb{D}   \, , \label{eq:dDdt-DR} \\
\partial_t\delta \chi _\p & = & - \nabla \cdot \mtb{D} - \mtb{f} \cdot \nabla  x_\p  \, ,  \label{eq:dchipdt-DR}
\end{eqnarray}
where we have defined
\begin{eqnarray}
\gamma_{\varepsilon} & \equiv & \left( 1 - \bar \varepsilon \right)
^{-1} \, , \\
\alpha^2 & \equiv & \left[ \, x_\p \left( 1 - x_\p
\right) \rho \right] \frac{1}{\rho} \left. \frac{\partial \beta }{
\partial x_\p } \right|_{\rho} = \frac{ c_s^2 \left( 1+\sigma \right)
\left( 1 - x_\p \right) } { \left[ 1 - \left( 1 + \sigma \right) x_\p
\right] } \, , \label{eq:alp2}
\end{eqnarray}
and where the speed of sound for an $N=1$~polytrope is given by $c_s^2
= 2 K \rho$. Now we assume that the perturbation variables behave as
plane waves, i.e. we introduce an $e^{i \left( \omega t - \mtb{k}
\cdot \mtb{r} \right) }$ dependence for all perturbations into
equations~(\ref{eq:dDdt-DR})--(\ref{eq:dchipdt-DR}). Here $\omega$ and
$\mtb{k}$ are the frequency and wave vector, respectively. The
characteristic polynomial of the resulting equations is then given by
\begin{equation}
\hat \omega ^4 - \left( \hat{\eta} ^2 \gamma_{\varepsilon} + 4
\gamma_{\varepsilon} ^{2} \hat \Omega^{2} \right) \hat \omega^2 + \hat{\eta} ^2
 \gamma_{\varepsilon}^3 \left( 2 \hat{\mtb{\Omega}} \cdot \hat{\mtb{k}} \right) ^2 = 0 \, ,
\label{eq:poly}
\end{equation}
where $\hat{\mtb{k}}$ is the unit wave vector and we have defined the
following dimensionless variables:
\begin{eqnarray}
\hat \Omega & \equiv & \frac{\Omega}{\sqrt{G \rho_c}} \, ,  \\
\hat \omega & \equiv & \frac{\omega}{\sqrt{G \rho_c}} \, ,  \\
\hat \eta   & \equiv & \frac{\alpha k}{\sqrt{G \rho_c}} \, .  \label{eq:eta}
\end{eqnarray}
The quantities $M$ and $R$ denote the mass and radius of a
non-rotating stellar model, respectively.  In the slow-rotation
approximation, we can assume that $\hat{\Omega} << \hat{\eta}$.
Equation~(\ref{eq:poly}) then have the following solutions [up to
$\mathcal{O}\left( \Omega^3\right)$]
\begin{eqnarray}
|\hat{\omega}_1 | & \simeq &  \sqrt{ \gamma_{\varepsilon}}  \, \hat{\eta}
                  + \frac{2 \, \gamma_{\varepsilon}^{3/2} }{ \hat{\eta} }
 \left[ \hat{\Omega}^2  - \left( \hat{\mtb{\Omega}} \cdot \hat{\mtb{k}} \right)^2\right]   \, , \label{eq:LAp} \\
|\hat{\omega}_2| & \simeq & \gamma_{\varepsilon} \, 2  \hat{\mtb{\Omega}} \cdot \hat{\mtb{k}} - \frac{4 \, \gamma_{\varepsilon} ^2 } { \hat{\eta}^2  }
\left[ \hat{\Omega}^2 - \left( \hat{\mtb{\Omega}} \cdot \hat{\mtb{k}} \right)^2 \right]
\hat{\mtb{\Omega}} \cdot \hat{\mtb{k}}
\simeq \gamma_{\varepsilon} \, 2  \hat{\mtb{\Omega}} \cdot \hat{\mtb{k}} \left\{ 1 - \frac{2 \, \gamma_{\varepsilon}  } { \hat{\eta}^2 }
\left[ \hat{\Omega}^2 - \left( \hat{\mtb{\Omega}} \cdot \hat{\mtb{k}} \right)^2 \right] \right\}   \, .  \label{eq:LAr}
\end{eqnarray}

As an estimate we consider a non-rotating background model with an
$N=1$~polytropic EoS. Using the definitions~(\ref{eq:alp2})
and~(\ref{eq:eta}) we have
\begin{equation}
\hat{\eta}^2 \simeq \frac{48}{\pi} \frac{ \left( 1+\sigma \right)
\left( 1 - x_\p \right) } { \left[ 1 - \left( 1 + \sigma \right) x_\p
\right] }\left( \frac{R}{\lambda} \right) ^2 \, ,   \label{eq:eta-bap}
\end{equation}
where $\lambda$ is the wavelength (related to the the wave vector
according to $k=2\pi/\lambda$). In addition, we have replaced the
speed of sound with its average value for an $N=1$~polytrope:
\begin{equation}
\langle c_s^2 \rangle = \frac{3}{\pi^2} \frac{G M}{R} \, .
\end{equation}

%------------------------------FIG. 7------------------------------------------%
\begin{figure}
\begin{center}
\includegraphics[height=75mm]{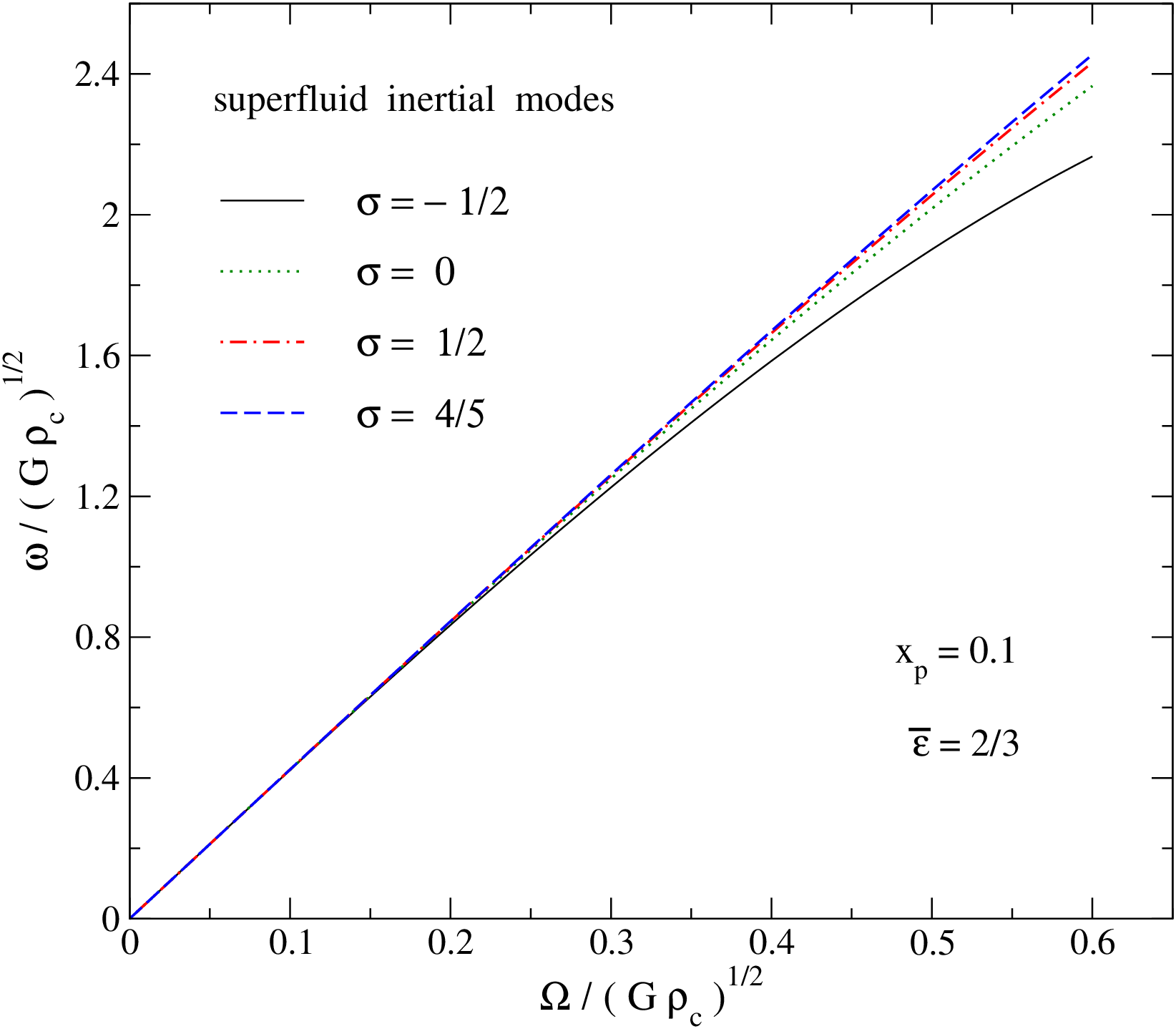}
\caption{This figure shows the inertial mode frequencies estimated
from the dispersion relation~(\ref{eq:LAp}). We consider the four
values of the symmetry energy term given in the legend. In
equation~(\ref{eq:LAp}), the correction term at $\mathcal{O}
(\Omega^3)$ leads to a dependence on $\sigma$ which is very similar to
the behaviour seen for the numerically determined mode-frequencies,
see figure~\ref{fig:rmodes} in the main text.
\label{fig:rmodes-LA}}
\end{center}
\end{figure}

If we now assume that the wavelength of the mode is of the same order
as the stellar radius, $\lambda=R$, equation~(\ref{eq:eta-bap}) for
$\hat{\eta}$ becomes
\begin{equation}
\hat{\eta}^2 \simeq \frac{48}{\pi} \frac{ \left( 1+\sigma \right) \left( 1 - x_\p
\right) } { \left[ 1 - \left( 1 + \sigma \right) x_\p \right] }\, .
\label{eq:eta-b}
\end{equation}
To get a qualitative picture, we consider parameter values $\bar
\varepsilon = 2 / 3$ and $ x_\p = 0.1$ and specify the angle between
$\hat{\mtb{\Omega}}$ and $\hat{\mtb{k}}$ to be $\theta = \pi/4$. In
figure~\ref{fig:rmodes-LA}, we show the positive frequency
$\hat{\omega}_\s$ of equation~(\ref{eq:LAp}) for different values of
$\sigma$. The $\mathcal{O} \left( \Omega ^ 3
\right)$ correction term leads to a behaviour that resembles the global mode results
shown in figure~\ref{fig:rmodes} in the main text.
%------------------------------------------------------------------------------%

%%%%%%%%%%%%%%%%%%%%%%%%%%%%%%%%%  BIBLIOGRAPHY  %%%%%%%%%%%%%%%%%%%%%%%%%%%%%%%
\nocite*
%\bibliographystyle{apsrev}
\bibliographystyle{mn2e}
%\bibliography{references}

%%%%%%%%%%%%%%%%%%%%%%%%%%%%%%%%%  LAST PAGE %%%%%%%%%%%%%%%%%%%%%%%%%%%%%%%
\label{lastpage}

%%%%%%%%%%%%%%%%%%%%%%%%%%%%%%%%%% END %%%%%%%%%%%%%%%%%%%%%%%%%%%%%%%%%%%%%%%%%%%%%%
\end{document}